\documentclass[useAMS,usenatbib]{mn2e}
\usepackage{amsmath}
\usepackage{amssymb}
\usepackage{graphicx}
\def\gsim{\;\lower4pt\hbox{${\buildrel\displaystyle >\over\sim}$}\;}
\def\lsim{\;\lower4pt\hbox{${\buildrel\displaystyle <\over\sim}$}\;}
\def\grls{\;\lower4pt\hbox{${\buildrel\displaystyle >\over <}$}\;}
\title[Two-Fluid Shock Flows of Galaxy Clusters]
{Self-Similar Shocks and Winds in Galaxy Clusters}
\author[Y.-Q. Lou, Y.-F. Jiang, C.-C. Jin ]{Yu-Qing Lou$^{1,2,3,}$
\thanks{Email: louyq@tsinghua.edu.cn; lou@oddjob.uchicago.edu},
Yan-Fei Jiang$^{1}$\thanks{jiangyanfei1986@gmail.com} and
Chi-Chuan Jin$^{1}$\thanks{jcc04@mails.tsinghua.edu.cn}\\
$^{1}$ Department of Physics and Tsinghua Centre for
Astrophysics (THCA), Tsinghua University, Beijing, 100084, China;\\
$^{2}$ Department of Astronomy and Astrophysics, the University of
Chicago, 5640 South Ellis Avenue, Chicago, IL 60637, USA;\\
$^{3}$ National Astronomical Observatories, Chinese Academy of
Sciences, A20, Datun Road, Beijing, 100021, China}
\begin{document}
\date{Accepted 2008 January 18. Received 2007 December 5;
in original form 2007 September 19} \maketitle
\begin{abstract}
A theoretical model framework of spherical symmetry is presented for
a composite astrophysical system of two polytropic fluids coupled
together by gravity to explore large-scale shocks and flow dynamics
in clusters of galaxies or in globular clusters. The existence of
such large-scale shocks in clusters of galaxies as inferred by
high-resolution X-ray and radio imaging observations implies
large-scale systematic flows that are beyond usual static models for
clusters of galaxies. Here, we explore self-similar two-fluid flow
solutions with shocks for a hot polytropic gas flow in a cluster of
galaxies in the presence of a massive dark matter (DM) flow after
the initiation of a gravitational core collapse or a central AGN
activity or a large-scale merging process. In particular, the
possibility of DM shocks or sharp jumps of mass density and of
velocity dispersion in dark matter halo is discussed and such DM
shocks might be detectable through gravitational lensing effects.
To examine various plausible scenarios for clusters of galaxies, we
describe three possible classes of shock flows within our model
framework for different types of temperature, density and flow speed
profiles. Depending upon sensible model parameters and shock
locations, the hot ICM and DM halo may have various combinations of
asymptotic behaviours of outflow, breeze, inflow, contraction or
static envelopes at large radii at a given time. We refer to
asymptotic outflows of hot ICM at large radii as the {\it galaxy
cluster wind}.
As a result of such galaxy cluster winds and simultaneous
contractions of DM halo during the course of galaxy cluster
evolution, there would be less hot ICM within clusters of galaxies
as compared to the average baryon fraction in the Universe.
Physically, it is then expected that such `missing baryons' with
lower temperatures reside in the periphery of galaxy clusters on
much larger scales. Based on our model analysis, we also predict a
limiting (the steepest) radial scaling form for mass density
profiles of $r^{-3}$ within clusters of galaxies.
\end{abstract}

\section{Introduction}

Extensive X-ray observations have revealed that almost completely
ionized hot gas medium permeates within clusters of galaxies with a
typical temperature range of $\sim 10^7-10^8$ K and a range of
typical electron number density $\sim 10^{-2}-10^{-4}\hbox{
cm}^{-3}$ (e.g., Cavaliere \& Fusco-Femiano 1978; Sarazin 1988;
Fabian 1994). Clusters of galaxies are largely gravitationally bound
systems on spatial scales of several Mpcs; together with the strong
evidence of high velocity dispersions of galaxies (e.g., $\sim
700-1000\hbox{ km s}^{-1}$), hot X-ray emitting thermal electron
gas, and gravitational lensing effects, we have realized the
presence of massive dark matter halo within clusters of galaxies.
Physical properties of galaxy clusters with static models of
spherical symmetry have been extensively studied in the past (e.g.,
Lea 1975; Cavaliere \& Fusco-Femiano 1976, 1978; Sarazin \& Bahcall
1977; Sarazin 1988; Fabian 1994; Carilli \& Taylor 2002; Voit 2005).

In the past several years, high-resolution X-ray imaging
observations have revealed density, temperature and pressure jumps
in the hot intracluster medium (ICM) within clusters of galaxies
(e.g., Fabian et al. 2003; Nulsen et al. 2005a; Nulsen et al. 2005b;
McNamara et al. 2005), indicating that these structures are likely
large-scale shocks rather than cold fronts (e.g., Sanders \& Fabian
2006). Active galactic nuclei (AGNs) (e.g., Nulsen et al. 2005a;
Nulsen et al. 2005b; McNamara et al. 2005) and merging galaxies
(e.g., Markevitch \& Vikhlinin 2001; Markevitch et al. 2002; Gabici
\& Blasi 2003; Markevitch et al. 2005) are proposed to be the
driving force and energy source of these large-scale shocks.
In addition, large-scale sound waves in clusters of galaxies have
been proposed by Sanders \& Fabian (2007) to explain the observed
quasi-concentric ripples in surface brightness of X-ray emissions.
Once shocks are identified in clusters of galaxies, there must be
large-scale flows involved. In other words, these clusters of
galaxies cannot be really static on large scales at least in the
spatial region where shocks are presumably identified. Relative to
the cluster centre, radial distances of shocks observed vary from
tens of kpcs (e.g., McNamara et al. 2005; Nulsen et al. 2005b) to
several Mpcs (e.g., A3667 in Rotteringer et al. 1997; A3376 in
Bagchi et al. 2006; galaxy clusters A786, A2255, A2256 in Ensslin et
al. 1998); at yet smaller radii around the centre, there could be
emerging shocks that may not be easily identified. We take the point
of view that these shocks are moving in hot ICM and we simply catch
them at different epochs of evolution. These shocks may all be born
somewhere around the central region and travel outwards to the
locations we observe at the present epoch. For those clusters of
galaxies without shock signatures, one possibility is that shocks
have occurred in the distant past and have disappeared after their
energies were dissipated during the process of propagation. So
large-scale shocks and flows may well be common phenomena in
clusters of galaxies.

On much smaller scales compared to those of clusters of galaxies,
X-rays have been also observed in globular clusters (e.g., Verbunt
et al. 1984) and these emissions are interpreted by some as
associated with flowing gas towards a black hole residing in the
centre of globular clusters (e.g., Silk \& Arons 1975). Moreover,
outflows of gas materials from globular clusters have also been
discussed in the literature (e.g., VandenBerg 1978). Fully relaxed
globular clusters can be well treated as spherically symmetric
(e.g., Harris \& Racine 1979). One may view the collection of stars
as one `fluid' and the tenuous gas as another fluid; these two
fluids are coupled together by gravity on large scales. Here, we
note globular clusters in passing and will mainly focus on
large-scale self-similar dynamics for clusters of galaxies.

With these two classes of astrophysical systems in mind, we develop
a theoretical model framework of spherical symmetry to study dynamic
behaviours of hot ICM and dark matter halo using the two-fluid
approximation (e.g., Lou 2005), where the two polytropic fluids are
coupled together by gravity. We should note that here the notion of
a polytropic fluid is fairly general in the sense of specific
entropy conservation along streamlines (Lou \& Cao 2007).

In the context of large-scale structure formation, extensive
numerical works have been carried out to simulate the formation of
galaxy clusters in the expanding universe, providing information for
the hot ICM and dark matter halo (see, e.g., Bertschinger 1998 for a
review on numerical simulations of structure formation in the
universe). Evrard (1990) and Thomas \& Couchman (1992) simulated
properties (such as the number density and temperature profiles) of
a hot gas in the presence of a dark matter halo. In Katz \& White
(1993) and Frenk et al. (1996), the radial cooling process is
simulated. In particular, Evrard et al. (1994) simulated the
formation of galaxies with two gravitationally coupled fluids
representing dark matter halo and baryon matter, which is similar in
essence to the approximation adopted in our semi-analytical model
for clusters of galaxies but on much larger scales.

Various self-similar solutions describing hydrodynamic processes of
a single self-gravitational isothermal or polytropic gas under
spherical symmetry have been investigated previously in contexts of
star formation (e.g., Larson 1969a; Larson 1969b; Shu 1977; Hunter
1977; Shu et al. 1987; Lou \& Shen 2004). Very recently, asymptotic
behaviours of novel quasi-static solutions in a single polytropic
gas sphere with self-gravity have been reported by Lou \& Wang
(2006, 2007) and was utilized to model rebound (MHD) shocks in
supernovae (Wang \& Lou 2007). For an astrophysical system of two
fluids coupled by gravity, we can systematically extend these
self-similar solutions, especially the new quasi-static solution,
which may be used to describe behaviours of hot ICM and dark matter
halo in clusters of galaxies. Except for the gravitational effect in
the Newtonian sense, nothing else is known about dark matter
particles at present. Using the coupled two-fluid model, we might be
able to learn physical properties of dark matter halo through
detectable diagnostics of hot ICM and of gravitational lensing
effects.

For clusters of galaxies, there is an outstanding problem of
`missing baryons'. Extensive X-ray observations have indicated that
the baryon mass fraction in clusters of galaxies is typically less
than the prediction of primordial nucleosynthesis (e.g., Ettori \&
Fabian 1999; Ettori 2003; He et al. 2005; McGaugh 2007). This
discrepancy becomes more difficult to reconcile in the cores of
galaxy clusters (e.g., Sand et al. 2003). The best fit of
cosmological parameters with tiny temperature fluctuations of the
cosmic microwave background (CMB) radiation and large-scale
structure clustering shows that relative to the critical mass
density $\rho_c$ in the universe, the mass density of baryon matter
is $\Omega_b=0.0224\pm 0.0009 h^{-2}_{100}$ and the total matter
density is $\Omega_m=0.135^{+0.008}_{-0.009}h^{-2}_{100}$,
where parameter $h_{100}$ is related to the Hubble constant $H_0$
by $H_0=100h_{100}$ km s$^{-1}$ Mpc$^{-1}$. Therefore, the mean
cosmic baryon mass fraction is
$f_b\equiv\Omega_b/\Omega_m=0.166^{+0.012}_{-0.013}$ (e.g., He et
al. 2005 and references therein).
While there are different methods in determining the $f_b$ value,
the cosmic baryon fraction $f_b$ is around 0.17 (e.g., McGaugh
2007). However in clusters of galaxies, the average gas (baryon)
fraction inferred by two methods are about $0.107^{+0.028}_{-0.019}$
and $0.111^{+0.069}_{-0.063}$ (e.g., Ettori 2003). Others estimated
that the baryon fraction $f_b$ observed in clusters of galaxies can
be lower than the cosmic baryon fraction by about $10\% - 20\%$ at
$z=0$ (e.g., He et al. 2005). Some even claimed that the value of
$f_b$ can be lowered by as much as 30\% (e.g., Ettori 2003). In
conclusion, the baryon fraction in most clusters of galaxies are
systematically lower than the average cosmic value $f_b$ except
those highest estimates for gas (baryon) mass fraction in some
clusters of galaxies (e.g., A426, A2142, RXJ1350; see Ettori 2003).
To resolve this important issue, the notion of Warm-Hot
Intergalactic Medium (WHIM) has been introduced (e.g., Cen \&
Ostriker 1999, 2006; Ettori 2003). In their opinion, the WHIM may
actually exist within clusters of galaxies to account for the mass
of `missing baryons', yet the WHIM cannot be detected at present
because it does not emit X-rays. These results show that a
significant fraction ($\sim 40\%-50\%$) of the baryon component
might be found in the form of WHIM in the temperature range of
$T\sim 10^{5-7}$ K (e.g., Cen \& Ostriker 2006).

As will be discussed in more details, this problem of `missing
baryons' in our model scenario is a natural consequence of galaxy
cluster winds, be it sustained or sporadic or be it stationary or
dynamic during the evolution of galaxy clusters. These so-called
`missing baryons' are blown away in the form of hot ICM and cool
down gradually with time; with relatively low temperatures, they
should mostly reside in the periphery of galaxy clusters and spread
out in space on much larger scales.
Meanwhile, the dark matter halo may contract within clusters of
galaxies in our model. Therefore the mass fraction of baryons $f_b$
(i.e., the mass ratio of total baryons to the total gravitational
mass inferred) would be lower than the initial value when a cluster
of galaxies was born and started to evolve.
The age of galaxy clusters is estimated to fall in the range of
$\sim 10^9\ -\ 10^{10}$ yr (e.g., Fabian 1994).
As galaxy cluster winds may have existed since galaxy clusters were
born, the timescale of galaxy cluster winds would be comparable to
or somewhat less than this estimate.
In Section 3, we show a few specific examples of numerical shock
flow solutions in our model and estimate the loss of baryons within
a timescale of $\sim 10^9$ yr.

As different behaviours of temperature profiles have been inferred
from X-ray observations of galaxy clusters (Markevitch 1996 and
Markevitch et al. 2005 for decreasing temperatures with increasing
radius; Peres et al. 1998 and Sanders \& Fabian 2006 for nearly
constant temperatures in several galaxy clusters; McNamara et al.
2005 and Blanton et al. 2001 for increasing temperatures with
increasing radius) and electron number densities are observed to fit
a power law fairly well (e.g., Peres et al. 1998; Nulsen et al.
2005b), we shall take the specific entropy conservation along
streamlines as the equation of state and see how well this may
account for the various observed profiles of thermodynamic
variables. By properly choosing model parameters in various regimes,
we can describe properties of galaxy clusters to a considerable
extent.

This paper is structured as follows. The background and motivation
of our model development is introduced in Section 1. Section 2
presents in order the basic formulation for the two-fluid model of
spherical symmetry involving two polytropic fluids, self-similar
transformation, asymptotic solutions at small and large $x$,
singular surfaces and sonic critical curves, and shock conditions.
In Section 3, we show numerical examples of quasi-static solutions
for three different situations. The major results are summarized in
Section 4. Finally we discuss our model results and numerical
solutions in Section 5. Certain mathematical details are contained
in Appendices A through G for the convenience of reference.

\section{Model Formulation}

As theoretical idealization and simplification, a cluster of
galaxies is approximated as fully relaxed or virialized and usually
modelled as a static equilibrium system in radial force balance with
spherical symmetry. However, large-scale shock features as observed
in clusters of galaxies reveal the presence of large-scale flows,
although these implied large-scale flows may not be directly
measurable at this stage. Our main motivation of this model analysis
is to provide a class of dynamic (rather than static or stationary)
models for clusters of galaxies with spherical symmetry. We hope to
understand a few basic aspects of this dynamic model framework. We
have several plausible processes in mind. First, the formation of
clusters of galaxies through large-scale gravitational collapse
involving dark matter and baryon matter. Secondly, activities of a
central AGN (involving accretions of baryon matter as well as dark
matter; e.g., Hu et al. 2004) onto supermassive black holes may give
rise to a quasi-spherically symmetric component of disturbances on
large scales which can evolve into shocks. Thirdly, merging
processes may reach a later phase of core confinement such that a
large-scale quasi-spherical symmetry may be a sensible
approximation; while releasing energy, it takes time for such a
dynamic system to relax and adjust itself.

\subsection{Self-Similar Equations for a Two-Fluid Model}

To study dynamic behaviours of visible baryon matter (such as X-ray
and radio emissions from the hot ICM) under the joint gravity of
both massive dark matter and baryon matter together, we adopt three
assumptions for the dark matter halo. First, self-interacting dark
matter particle models have been proposed earlier by some
researchers (e.g., Carlson et al. 1992; Machacek 1994; Spergel \&
Steinhardt 2000) to solve the problems encountered by cold dark
matter models.
Furthermore, properties of collisional dark matter particles (e.g.,
Ostriker 2000; Hu \& Lou 2007) and fluid dark matter (e.g., Peebles
2000; Subramanian 2000; Moore et al. 2000; Hennawi \& Ostriker 2002;
Lou 2005; Hu et al. 2006) were proposed as an alternative approach
to probe DM dynamics. On large scales, one may view high velocity
dispersions ($\sim 700-1000\hbox{ km s}^{-1}$) of DM particles to
produce an effective pressure against gravity as described by the
Jeans equation (e.g., Binney \& Tremaine 1987). In particular,
Evrard et al. (1994) numerically simulated formation of galaxies
using a model consisting of two gravitationally coupled fluids
representing dark matter and baryon matter. While a distribution
function approach can be applied to study properties of galaxy
clusters, we model a dark matter halo in a `fluid' approximation to
simplify the mathematical treatment. On large scales and without
resonances, we should be able to understand various dynamic
behaviours of hot intracluster medium (ICM) and the dark matter halo
in this two-fluid approximation (Lou 2005).
Secondly, we assume the two-fluid system of galaxy clusters to be
spherically symmetric with a common centre and without rotation for
simplicity. On much smaller scales, this simplification would be a
very good approximation for globular clusters containing millions of
stars and gas. Thirdly, dark matter interacts with hot ICM only
through gravity. Based on the above assumptions, we readily write
out a set of coupled nonlinear partial differential equations to
describe the two-fluid flow system with hot ICM and dark matter halo
coupled by gravity. In spherical polar coordinates $(r,\ \theta,\
\phi)$, the equation for mass conservation is described by
\begin{eqnarray}
\frac{\partial M_i}{\partial t}+u_i\frac{\partial M_i}{\partial r}=0
\qquad \mbox{  and  }\qquad \frac{\partial M_i}{\partial r}=4\pi
r^2\rho_i\ , \label{eq01}
\end{eqnarray}
where $r$ is radius and $t$ is time; subscripts $i=1,\ 2$ stand for
dark matter halo (fluid $1$) and hot ICM (fluid $2$), respectively.
For simplicity, all variables with a subscript $i$ denote
associations with fluid $i$. When referring to a specific fluid $i$,
we would write out subscripts $1$ or $2$ explicitly. Here,
$M_i(r,t)$, $\rho_i(r,t)$, and $u_i(r,t)$ are respectively the
enclosed mass, the mass density, and the radial flow speed at radius
$r$ and time $t$ for fluid $i$. Another familiar form equivalent to
the above continuity equation (\ref{eq01}) is
\begin{eqnarray}
\frac{\partial\rho_i}{\partial
t}+\frac{1}{r^2}\frac{\partial}{\partial r}(r^2\rho_i u_i)=0\ .
\label{eq02}
\end{eqnarray}
The Euler radial momentum equation
is
\begin{eqnarray}
\frac{\partial u_i}{\partial t}+u_i\frac{\partial u_i}{\partial
r}=-\frac{1}{\rho_i}\frac{\partial P_i}{\partial
r}-\frac{G(M_1+M_2)}{r^2}\ ,\label{eqforce}
\end{eqnarray}
where $P_i(r,t)$ is the pressure at radius $r$ and time $t$ for
fluid $i$; $M_1$ and $M_2$ are the enclosed masses of fluid $1$ and
fluid $2$, respectively; $G=6.67\times10^{-11}\
\hbox{kg}^{-1}\hbox{m}^3\hbox{s}^{-2}$ is the gravitational
constant.\footnote{In Lou \& Wang (2006), there is a typo in the
unit of $G$; it should be cm$^3$ instead of cm$^{-3}$.}
The coupling effect of the two fluids is explicitly contained in
the gravity term in equation (\ref{eqforce}). Finally, we take the
conservation equation of specific `entropy' along streamlines as
the equation of state (e.g., Fatuzzo et al. 2004; Wang \& Lou
2007; Lou \& Cao 2007)
\begin{eqnarray}
\left(\frac{\partial}{\partial t}
+u_i\frac{\partial}{\partial r}\right)
\left(\frac{P_i}{\rho_i^{\gamma_i}}\right)=0\ ,
 \label{state}
\end{eqnarray}
where $\gamma_i$ is the polytropic index for fluid $i$. In
general, polytropic indices $\gamma_1$ and $\gamma_2$ are allowed
to be different.

Equations $(\ref{eq01})-(\ref{state})$ form a set of nonlinear
partial differential equations and contain an important subset of
nonlinear self-similar solutions with or without shocks. We now
introduce a set of self-similar transformation for the two
polytropic fluids (e.g., Suto \& Silk 1988; Lou \& Wang 2006)
below
\begin{eqnarray}
\lefteqn{x_i=\frac{r}{K_i^{1/2}t^{n_i}}\ ,\ \
\rho_i=\frac{\alpha_i(x_i)}{4\pi Gt^2}\ ,\ \
u_i=K_i^{1/2}t^{n_i-1}v_i(x_i)\ ,}\ \nonumber\\
&&\!\!\!\!\!\!\!\!\!\!\!\!\!\!\!\! P_i=\frac{K_it^{2n_i-4}}{4\pi
G}\beta_i(x_i)\ ,\qquad
M_i=\frac{K_i^{3/2}t^{3n_i-2}}{(3n_i-2)G}m_i(x_i)\
,\label{transformation}
\end{eqnarray}
where $x_i$ is the independent dimensionless similarity variable for
fluid $i$ ($x_1$ and $x_2$ are actually related to each other);
$K_i$ is a scale parameter of self-similar transformation for fluid
$i$; $n_i$ is another index parameter of self-similar transformation
for fluid $i$ noted above; $\alpha_i(x_i)$ is the reduced mass
density for fluid $i$; $v_i(x_i)$ is the reduced radial flow speed
for fluid $i$; $\beta_i(x_i)$ is the reduced pressure for fluid $i$;
and $m_i(x_i)$ is the reduced enclosed mass for fluid $i$. All these
reduced variables are functions of independent variable $x_i$ only.
In principle, parameter $n_1$ does not need to be equal $n_2$. But
equation (\ref{eqforce}) contains factors like $t^{n_1}$ and
$t^{n_2}$ at the same time after the self-similar transformation.
Therefore in order to obtain the self-similar dimensionless ordinary
differential equations (ODEs) in terms of $x_i$ without involving
explicit temporal factors of time $t$, we simply require $n_1=n_2=n$
as a single parameter and shall not distinguish the two from now on.
With the specific entropy conservation (\ref{state}) along
streamlines, it is not necessary to require $n=2-\gamma_i$ here
(e.g., Yahil 1983; Suto \& Silk 1988; Lou \& Wang 2006, 2007). In
fact with $1<\gamma_i<2$, we shall explore three possible situations
of $2/3<n<1$, $n=1$ and $n>1$, respectively. We note that
self-similar processes of galaxy cluster evolution and of cooling
waves in galaxy clusters have also been studied previously (e.g.,
Bertschinger 1989; Jain \& Bertschinger 1996). Here, cooling waves
refer to a self-similar expansion of cooling flow region. While the
cooling region expands, the hot ICM itself does not move out.

By performing self-similar transformation (\ref{transformation}),
the reduced enclosed mass $m_i(x_i)$ can be expressed as (e.g.,
Suto \& Silk 1988; Lou \& Wang 2006; Lou \& Cao 2007)
\begin{eqnarray}
m_i(x_i)=\alpha_ix_i^2(nx_i-v_i)\ .  \label{mass}
\end{eqnarray}
Since the enclosed mass $M_i>0$, we should require $m_i>0$ and
$nx_i>v_i$ for $n>2/3$, while for $n<2/3$, we require $m_i<0$ and
$nx_i<v_i$; the latter is generally impossible for semi-complete
solutions in the range of $0^{+}<x<+\infty$. Using equation
(\ref{mass}) and self-similar transformation
(\ref{transformation}), specific entropy conservation
(\ref{state}) along streamlines leads to
\begin{eqnarray}
\frac{\beta_i(x_i)}{\alpha_i^{\gamma_i}(x_i)}
=C_{0,i}m_i^{q_i}(x_i)\ ,
\end{eqnarray}
where $q_i\equiv 2(n+\gamma_i-2)/(3n-2)$ is a naturally emerged
index parameter and $C_{0,i}$ is an integration constant for each
fluid $i$. For $\gamma_i\neq 4/3$, we can always effectively combine
the two coefficients $C_{0,i}$ and $K_i$ into a new single constant
coefficient,
corresponding to a coefficient rescaling in self-similar
transformation (\ref{transformation}). The case of a single fluid
with $\gamma=4/3$ is separately considered by Lou \& Cao (2007;
see also Goldreich \& Weber 1980 and Yahil 1983). It then suffices
to consider the equation of state in the form of
\begin{eqnarray}
\frac{\beta_i(x_i)}{\alpha_i^{\gamma_i}(x_i)}=m_i^{q_i}(x_i)\
\label{stateless}
\end{eqnarray}
for fluid $i$ with coefficients $C_{0,i}$ being absorbed without
loss of generality. Apparently, there are two linearly related
independent variables $x_1$ and $x_2$ respectively for the two
coupled fluids under consideration. We introduce a convenient
ratio $\kappa\equiv (K_1/K_2)^{1/2}$ such that $\kappa=x_2/x_1$.
We can then express all dependent variables as functions of $x_1$
only. From now on, we shall rewrite $x_1$ as $x$ for simplicity
and thus $x_2=\kappa x$. Now mass and momentum conservation
equations (\ref{eq02}) and (\ref{eqforce}) can be
straightforwardly cast into the following dimensionless forms of
four ODEs, namely
\begin{eqnarray}
(nx-v_1)\frac{d\alpha_1}{dx}-\alpha_1\frac{dv_1}{dx}
=-2\frac{(x-v_1)}{x}\alpha_1\ , \label{Da1}
\end{eqnarray}
\begin{eqnarray}
\lefteqn{(n-1)v_1-(nx-v_1)\frac{dv_1}{dx}=}\nonumber\\
&&{}-\gamma_1\alpha_1^{q_1+\gamma_1-2}x^{2q_1}(nx-v_1)^{q_1}
\frac{d\alpha_1}{dx}\nonumber\\
&&{}-q_1\alpha_1^{q_1+\gamma_1-1}x^{2q_1}
(nx-v_1)^{q_1-1}(3n-2)\nonumber\\
&&{}-\frac{\alpha_1(nx-v_1)}{(3n-2)}-\frac{\alpha_2(\kappa
nx-v_2)}{(3n-2)\kappa}\ , \label{dv1}
\end{eqnarray}
\begin{eqnarray}
(\kappa nx-v_2)\frac{d\alpha_2}{dx}-\alpha_2\frac{dv_2}{dx}
=-2\frac{(\kappa x-v_2)}{x}\alpha_2\ ,\label{ga2}
\end{eqnarray}
\begin{eqnarray}
\lefteqn{(n-1)v_2-\frac{(\kappa nx-v_2)}{\kappa}\frac{dv_2}{dx}=}\nonumber\\
&&{}-\frac{\gamma_2}{\kappa}\alpha_2^{q_2+\gamma_2-2}\kappa^{2q_2}
x^{2q_2}(\kappa nx-v_2)^{q_2}\frac{d\alpha_2}{dx}\nonumber\\
&&{}-q_2\alpha_2^{q_2+\gamma_2-1}\kappa^{2q_2}
x^{2q_2}(\kappa nx-v_2)^{q_2-1}(3n-2)\nonumber\\
&&{}-\frac{\alpha_2(\kappa nx-v_2)}{(3n-2)}
-\frac{\kappa\alpha_1(nx-v_1)}{(3n-2)}\ .\label{gv2}
\end{eqnarray}
After rearrangement and algebraic manipulations of above four
equations (\ref{Da1})$-$(\ref{gv2}), the reduced radial flow speeds
$v_1(x)$ and $v_2(x)$ and the reduced mass densities $\alpha_1(x)$
and $\alpha_2(x)$ are then determined equivalently by four coupled
first-order nonlinear ODEs shown below.
\begin{eqnarray}\label{da1}
\frac{d\alpha_1(x)}{dx}=\frac{{\cal A}_1(x)}{{\cal D}_1(x)}\ ,
\label{a1}
\end{eqnarray}
\begin{eqnarray}\label{dv1}
\frac{dv_1(x)}{dx}=\frac{{\cal V}_1(x)}{{\cal D}_1(x)}\ ,
\label{dv1}
\end{eqnarray}
\begin{eqnarray}\label{da2}
\frac{d\alpha_2(x)}{dx}=\frac{{\cal A}_2(x)}{{\cal D}_2(x)}\ ,
\label{da2}
\end{eqnarray}
\begin{eqnarray}\label{v2}
\frac{dv_2(x)}{dx}=\frac{{\cal V}_2(x)}{{\cal D}_2(x)}\ ,
\end{eqnarray}
where explicit expressions of denominators ${\cal D}_1(x)$ and
${\cal D}_2(x)$ and numerators ${\cal V}_1(x)$ and ${\cal V}_2(x)$
are shown in Appendix A. This set of nonlinear ODEs contains
various types of asymptotic solutions at large and small $x$. We
mainly focus on two of them, one at large $x$ and the other at
small $x$.

\subsection{Tests of Our Model Formulation}\label{test of the model}

Our model formulation here is sufficiently general and can be
readily reduced to various known formulations through various
paths of reduction. We indicate these different paths of reduction
below to test and confirm the robustness of our approach.

For $n=1$ and $\gamma_1=\gamma_2=1$, the formulation here reduces to
a spherical composite system of two isothermal fluids coupled by
gravity as explored by Lou (2005). Static and dynamic models may be
constructed in that relatively simple theoretical framework in
contexts of clusters of galaxies or of globular clusters but on much
smaller scales.

By setting physical variables of one of the two fluids to vanish,
the formulation here reduces to that of a single spherical system
for a more general polytropic fluid under self-gravity (e.g.,
Fatuzzo et al. 2004; Lou \& Cao 2007; Wang \& Lou 2007 in
preparation).

If we further set $n+\gamma=2$ for a single general polytropic gas
under self-gravity, our formulation here reduces to that of a
conventional polytropic gas with a constant specific entropy
distribution in time and space (Goldreich \& Weber 1980; Yahil 1984;
Suto \& Silk 1988; Lou \& Gao 2006; Lou \& Wang 2006, 2007).

The single isothermal fluid of spherical symmetry corresponds to
$n=1$ and $\gamma=1$ in our formulation. This problem has been
extensively explored in the literature (e.g., Larson 1969a, b;
Penston 1969a, b; Shu 1977; Whitworth \& Summers 1985; Hunter 1977,
1986; Tsai \& Hsu 1995; Shu et al. 2002; Lou \& Shen 2004; Shen \&
Lou 2004; Fatuzzo et al. 2004; Bian \& Lou 2005; Yu \& Lou 2005; Yu
et al. 2006).

Our model is formulated for two gravitationally
coupled fluids which are polytropic in the more general sense (see
equation \ref{state}); this holds the key difference in reference
to previous single fluid model under self-gravity with various
equations of state. We will show in this section by examples that
the solutions similar to the well-known solutions in the single
fluid framework (such as the static solution, the central
free-fall solution (Shu 1977) and the Larson$-$Penston type
solution (Larson 1969a, b; Penston 1969a, b)) can also be derived
within our model framework.

The static solution in which both flow velocities of the hot gas and
dark matter vanish throughout the entire space can be found in our
model framework. This is an exact global solution similar to the
static solutions with central divergence (i.e., singular isothermal
sphere (SIS) and singular polytropic sphere (SPS)) in single fluid
model framework (e.g., Shu 1977; Cheng 1978; Lou \& Shen 2004; Lou
\& Wang 2006; Lou \& Cao 2007). In our two-fluid model framework,
the global static solution (i.e., singular double polytropic spheres
(SDPS)) is simply
\begin{eqnarray}
v_1=v_2=0\ ,\ \ \alpha_1=A_1x^{-2/n}\ ,\ \ \alpha_2=A_2x^{-2/n}\
.\label{SPS}
\end{eqnarray}
Here, the two positive density coefficients $A_1$ and $A_2$ for
two static fluids are readily determined by the following pair of
equations
\begin{eqnarray}
\frac{n(A_1+A_2)}{2(3n-2)}=(2-n)n^{q_1-1}
A_1^{q_1+\gamma_1-1}\ ,\nonumber\\
\frac{n(A_1+A_2)}{2(3n-2)}=\kappa^{3q_2-2}
(2-n)n^{q_2-1}A_2^{q_2+\gamma_2-1}\ ,\label{staticdensity}
\end{eqnarray}
where $q_i\equiv 2(n+\gamma_i-2)/(3n-2)$ and the ratio $\kappa$ for
the two fluids breaks the symmetry of the above two relations; this
symmetry would be explicit for $\kappa=1$. For physical solutions
with positive $A_1$ and $A_2$, it is necessary to require $2/3<n<2$.
In other words, both power-law density scalings fall between
$x^{-1}$ and $x^{-3}$. For a set of four specified parameters ($n,\
\gamma_1,\ \gamma_2,\ \kappa$), equation (\ref{staticdensity}) does
possess sensible real solutions for both coefficients $A_1>0$ and
$A_2>0$ and our model then gives a singular static solution for both
polytropic fluids simultaneously, i.e., two gravity coupled singular
polytropic spheres (SPSs) with divergence of mass densities as
$x\rightarrow 0^{+}$. In astrophysical applications, we need to
introduce a proper central cutoff.

The LP asymptotic solution and Shu's central free-fall asymptotic
solution were constructed for a single isothermal sphere; in order
to get analogous asymptotic solutions in our two-fluid model, we set
scaling index parameter $n=1$ (defined in self-similar
transformation equation (\ref{transformation})) and the polytropic
indices of both fluids $\gamma_1=1$ and $\gamma_2=1$. This is a
self-gravitating system of two coupled singular isothermal spheres
(Lou 2005).

For two coupled isothermal gas spheres with $\kappa=1$, the
generalized version of the Einstein-de Sitter solution (Whitworth \&
Summers 1985; Shu et al. 2002; Lou \& Shen 2004; Lou \& Zhai 2007 in
preparation) in our model is
\begin{eqnarray}\label{EdSsolution}
v_1=v_2={2x}/{3}\ ,\ \ \ \alpha_1=\alpha_2={1}/{3}\ .
\end{eqnarray}
This isothermal solution is an exact global solution. For
$\kappa\neq 1$, this kind of solutions does not exist.

For two coupled conventional polytropic spheres with
$n+\gamma_1=n+\gamma_2=2$ and $\kappa=1$, the generalized version of
the Einstein-de Sitter solution of our model is still given by
equation (\ref{EdSsolution}) (Lou \& Wang 2006; Wang \& Lou 2007;
Lou \& Cao 2007).


The asymptotic behaviour when $x\rightarrow 0^{+}$ of the solution
similar to Shu's central free-fall asymptotic solution in our model
framework is
\begin{eqnarray}\label{vi}
v_1\rightarrow\mathcal{H}_1x^{-{1}/{2}}\ ,\ \ \ \ \ \
v_2\rightarrow\mathcal{H}_2x^{-{1}/{2}}\ ,
\end{eqnarray}
\begin{eqnarray}\label{alphai}
\alpha_1\rightarrow\mathcal{G}_1x^{-{3}/{2}}\ ,\ \ \ \ \ \
\alpha_2\rightarrow\mathcal{G}_2x^{-{3}/{2}}\ .
\end{eqnarray}
Here, the two relations among the four coefficients $\mathcal{H}_1$,
$\mathcal{H}_2$, $\mathcal{G}_1>0$ and $\mathcal{G}_2>0$ are
\begin{eqnarray}
\mathcal{H}_1^2+2\mathcal{H}_1\mathcal{G}_1
+2\frac{\mathcal{H}_2\mathcal{G}_2}{\kappa}=0\
,
\end{eqnarray}
\begin{eqnarray}
\mathcal{H}_2^2+2\kappa\mathcal{H}_2\mathcal{G}_2
+2\kappa^2\mathcal{H}_1\mathcal{G}_1=0\
.
\end{eqnarray}
Therefore, there are only two free coefficients for this kind of
central free-fall asymptotic solutions as $x\rightarrow 0^{+}$. The
mass at the centre for fluid $1$ are determined by
$m_1(0)=-\mathcal{H}_1\mathcal{G}_1$ and the mass at the centre for
fluid $2$ are determined by
$m_2(0)=-\kappa^2\mathcal{H}_2\mathcal{G}_2$. Therefore,
$\mathcal{H}_1$ and $\mathcal{H}_2$ are both negative in equation
(\ref{vi}), corresponding to central free-falls of both fluids.

For a polytropic single fluid model, quasi-static asymptotic
solutions have been constructed by Lou \& Wang (2006). In the next
subsection, we will also construct similar quasi-static asymptotic
solutions in our two-fluid model and apply the model to the
large-scale dynamics of hot gas and dark matter in clusters of
galaxies in this paper.

\subsection{Construction of Two-Fluid Quasi-Static Solutions}

As the sound speed of hot ICM in clusters of galaxies is of the
order of $\sim 1000\hbox{ km s}^{-1}$ and may decrease towards the
centre (such as Abell 2052 in Blanton et al. 2001, and MS0735.6+7421
in McNamara et al. 2005 and in Cavaliere \& Fusco-Femiano 1978)
while the observed radial flow speed of hot ICM close to the centre
is not large, we then construct the quasi-static solution from
equations (\ref{da1})$-$(\ref{v2}) in the regime of small $x$ (i.e.,
parallel to the quasi-static solution of a single fluid of Lou \&
Wang 2006 and Wang \& Lou 2007).\footnote{Yahil \& Ostriker (1973)
discussed a steady outflow of gas from galaxy clusters. Their steady
wind results show that the gas velocity towards the centre is also
very small (see their Fig. 3).}

Now we take the static SPS solution (\ref{SPS}) and
(\ref{staticdensity})
as given in the subsection \ref{test of the model} as the
leading-order term of the asymptotic quasi-static solution as
$x\rightarrow 0^{+}$. Expanding to the second order of this series,
the quasi-static solution for two gravity coupled polytropic fluids
then takes the form of
\begin{eqnarray}
v_1&\rightarrow& L_1x^{k}+\ldots\ ,\nonumber\\
v_2&\rightarrow& L_2x^{k}+\ldots\ ,\label{quasiv}
\end{eqnarray}
\begin{eqnarray}
\alpha_1&\rightarrow& A_1x^{-2/n}+N_1x^{R}+\dots\ ,\nonumber\\
\alpha_2&\rightarrow& A_2x^{-2/n}+N_2x^{R}+\dots\ ,\label{quasia}
\end{eqnarray}
where $R\equiv k-1-2/n$ and index parameter\footnote{The index
parameter $k$ is taken to be the same for both fluids.} $k$ is
determined by the following quartic algebraic equation
\begin{eqnarray}
C_{k,1}\ k^4+C_{k,2}\ k^3+C_{k,3}\ k^2+C_{k,4}\ k+C_{k,5}=0\
.\label{k}
\end{eqnarray}
Explicit expressions of the five coefficients $(C_{k,1},\ C_{k,2},\
C_{k,3},$ $C_{k,4},\ C_{k,5})$ can be found in Appendix B. In
general, the roots of algebraic quartic equation (\ref{k}) can be
complex. For the higher order terms with respect to static SPS
solution, we should require $Re(k)>1$ (in static SPS solution
(\ref{SPS}), $v_1=v_2=0x$ is regarded as the zeroth order); and the
two coefficients $L_1$ and $L_2$ are two free parameters with the
other two coefficients $N_1$ and $N_2$ readily determined by
\begin{eqnarray}\label{Ni}
N_1=\frac{[2(n-1)+nk]A_1L_1}{(k-1)n^2}\ ,\nonumber\\
N_2=\frac{[2(n-1)+nk]A_2L_2}{\kappa (k-1)n^2}\ .
\end{eqnarray}
For a complex index parameter $k$, please refer to Appendix C for
quasi-static solutions with asymptotic oscillations in the regime of
small $x$ (see also Lou \& Wang 2006 for quasi-static solutions with
asymptotic oscillatory behaviours).

\subsection{Asymptotic Solutions at Large $x$}
\label{asymptotic solutions at large x}

In this section, we shall derive asymptotic behaviours of radial
flow speed and mass density of the two fluids from equations
(\ref{da1})$-$(\ref{v2}) as $x\rightarrow +\infty$.

The asymptotic series solution at large $x$ takes the form of
\begin{eqnarray}
\alpha_1(x)\rightarrow E_1x^{-2/n}+I_1x^{-3/n}+\dots\ ,
\label{largea1}
\end{eqnarray}
\begin{eqnarray}
\alpha_2(x)\rightarrow E_2x^{-2/n}+I_2x^{-3/n}+\ldots\ ,
\label{largea2}
\end{eqnarray}
\begin{eqnarray}
v_1(x)\rightarrow H_1x^{-1/n+1}+G_1x^{-2/n+1}+\ldots\ ,
\label{largev1}
\end{eqnarray}
\begin{eqnarray}
v_2(x)\rightarrow H_2x^{-1/n+1}+G_2x^{-2/n+1}+\ldots\ ,
\label{largev2}
\end{eqnarray}
where $2/3<n<2$ is required and the four coefficients $E_1$, $E_2$,
$H_1$ and $H_2$ are fairly arbitrary, while the other four
coefficients $I_1$, $I_2$, $G_1$ and $G_2$ can be expressed in terms
of these four arbitrary coefficients (details of these expressions
are contained in Appendix D).

For asymptotic radial flow speed solutions (\ref{largev1}) and
(\ref{largev2}), the flow velocities diverge as $x\rightarrow
+\infty$ for $n>1$ unless $H_1=H_2=0$. For a real astrophysical
system, its size is finite; and we may need to introduce a spatial
cutoff at a given time in order to make use of these solutions for
$n>1$ with $H_1\neq 0$ and $H_2\neq 0$. For example, the typical
size of a galaxy cluster is of the order of several to ten Mpcs.

We are certainly interested in finite asymptotic solutions
(\ref{largev1}) and (\ref{largev2}) with $n=1$ and $2/3<n<1$.
While for $n>1$, it is possible to set $H_1=H_2=0$, and the
asymptotic solutions finite at large $x$ become
\begin{eqnarray}
\alpha_1(x)\rightarrow E_1x^{-2/n}+F_1x^{-4/n+1}+\ldots\
,\label{largea1_0}
\end{eqnarray}
\begin{eqnarray}
\alpha_2(x)\rightarrow E_2x^{-2/n}+F_2x^{-4/n+1}+\ldots\
,\label{largea2_0}
\end{eqnarray}
\begin{eqnarray}
v_1(x)\rightarrow G_1x^{-2/n+1}+D_1x^{-4/n+2}+\ldots\ ,
\label{largev12}
\end{eqnarray}
\begin{eqnarray}
v_2(x)\rightarrow G_2x^{-2/n+1}+D_2x^{-4/n+2}+\ldots\ ,
\label{largev22}
\end{eqnarray}
where $E_1$ and $E_2$ are two fairly arbitrary constants and the
condition $n<2$ is still required; $G_1$ and $G_2$ are still
defined in Appendix D. $F_1$, $F_2$, $D_1$, and $D_2$ are four
constant coefficients of next order expansion terms; these
coefficients are determined by specified values of $E_1$ and $E_2$
(further details of these coefficient expressions can be found in
Appendix D). In principle, we can carry out this series expansion
for large $x$ to the desired order if needed.

\subsection{Singular Surfaces and\\
\qquad\ Sonic Critical Curves}

The singular surface and sonic critical curve in a single gas flow
have been analyzed in details by previous authors (e.g., Suto \&
Silk 1988; Whitworth \& Summers 1985; Lou \& Wang 2006; Wang \& Lou
2007). For a single gas flow of spherical symmetry, there are smooth
solutions going across the sonic critical curve analytically (e.g.,
Suto \& Silk 1988; Whitworth \& Summers 1985; Lou \& Shen 2004) or
with shocks (e.g., Tsai  \& Hsu 1995; Shu et al. 2000; Bian \& Lou
2005; Lou \& Wang 2006; Yu et al. 2006; Wang \& Lou 2007; Lou \& Cao
2007). In our two-fluid model, each fluid component has its own
singular surface and sonic critical curve; they are not the same for
the two fluids in general and should be treated separately.

The singular surfaces are a set of points when ${\cal D}_1(x)=0$
for fluid $1$ by definition (\ref{D1}) and ${\cal D}_2(x)=0$ for
fluid $2$ by definition (\ref{D2}). Physically, when the travel
speed of disturbances relative to the flow speed is equal to the
local sound speed in a fluid, we encounter a singular surface. As
there are two fluids in our model, there are two singular surfaces
possible for a given set of relevant parameters. The singular
surface $\{x,\ v_1,\ \alpha_1\}$ of fluid $1$ as defined by ${\cal
D}_1(x)=0$ is then given the following equation
\begin{eqnarray}
(nx-v_1)^2-\gamma_1\alpha_1^{q_1+\gamma_1-1}
x^{2q_1}(nx-v_1)^{q_1}=0\
,
\end{eqnarray}
and the singular surface $\{x,\ v_2,\ \alpha_2\}$ of fluid $2$ as
defined by ${\cal D}_2(x)=0$ is given by a similar equation
\begin{eqnarray}
(\kappa nx-v_2)^2-\kappa^{2q_2}\gamma_2\alpha_2^{q_2+\gamma_2-1}
x^{2q_2}(\kappa nx-v_2)^{q_2}=0\ .
\end{eqnarray}

Mathematically, sonic critical curves are characteristic lines on
the singular surface with the numerators and denominators of
self-similar nonlinear ODEs (\ref{a1}) and (\ref{dv1}), and
(\ref{da2}) and (\ref{v2}) separately being zero simultaneously. The
intersection of such two surfaces in $\{x\ ,v_i\ ,\alpha_i\}$ space
gives the sonic critical curve in each fluid $i$. As noted above,
when the travel speed of disturbances relative to the local flow
speed of a specific fluid is equal to the local sound speed, a
singularity arises. If one wants to get continuous flow solutions
throughout the entire range of $x$, such solutions must pass across
the singular surface via the sonic critical curves.

In our two-fluid model, as ${\cal V}_1(x)$ and ${\cal A}_1(x)$
contain $\alpha_2$ and $v_2$, we need proper values of $\alpha_2$
and $v_2$ in order to determine the sonic critical curve of fluid
$1$; in reciprocal and in parallel, as ${\cal V}_2(x)$ and ${\cal
A}_2(x)$ contain $\alpha_1$ and $v_1$, we need proper values of
$\alpha_1$ and $v_1$ in order to determine the sonic critical
curve of fluid $2$. This is a major yet expected difference as
compared with the case of a single fluid. Therefore in principle,
for a given position of $x$, there is a critical point (a point on
the critical curve) for fluid $i$ once a set of ($\alpha_i$,
$v_i$) at that $x$ of the other fluid is given. As the two fluids
do not have their critical curves at the same $x$ in general,
there are only two eigendirections for a given set of parameters,
which is similar to the case of a single fluid model (Lou \& Wang
2006; Wang \& Lou 2007). The equations and specific procedure to
determine the eigendirections can be found in Appendix E.

In reference to our quasi-static solution for small $x$, the
condition for the corresponding singular surface ${\cal D}_1(x)=0$
of fluid $1$ has the asymptotic behaviour when
$x\rightarrow0^{+}$:
\begin{eqnarray}
nx&=&\Big(\gamma_1\alpha_1^{q_1+\gamma_1-1}
x^{2q_1}\Big)^{1/(2-q_1)}+v_1\nonumber\\
&\rightarrow &\Big[\gamma_1
A_1^{q_1+\gamma_1-1}\Big]^{1/(2-q_1)}x^{1-2/[n(2-q_1)]}\ .
\end{eqnarray}

For ${\cal D}_2(x)=0$, the asymptotic behaviour of the corresponding
singular surface for fluid $2$ when $x\rightarrow0^{+}$ is
\begin{eqnarray}
\kappa nx&=&\Big(\kappa^{2q_2}\gamma_2\alpha_2^{q_2+\gamma_2-1}
x^{2q_2}\Big)^{1/(2-q_2)}+v_2\nonumber\\
&\rightarrow&\Big[\gamma_2\kappa^{2q_2}
A_2^{q_2+\gamma_2-1}\Big]^{1/(2-q_2)}
x^{1-2/[n(2-q_2)]}\ .
\end{eqnarray}
Therefore if $1-2/[n(2-q_i)]>0$, the corresponding singular
surface of fluid $i$ passes through the origin point ($x=0$ and
$v_i=0$); otherwise, it cannot pass the zero point.

When $x\rightarrow +\infty$, the asymptotic behaviour of
$\alpha_i(x)$ is (see asymptotic solutions (\ref{largea1}),
(\ref{largea2}), (\ref{largea1_0}) and (\ref{largea2_0}))
characterized by $\alpha_i\rightarrow E_ix^{-2/n}+\ldots$. Then
for this same limit of large $x$, the singular surface condition
${\cal D}_1(x)=0$ becomes
\begin{eqnarray}
v_1&=&nx-\Big(\gamma_1\alpha_1^{q_1+\gamma_1-1}x^{2q_1}\Big)^{1/(2-q_1)}\nonumber\\
&\rightarrow&nx-\Big(\gamma_1E_1^{q_1+\gamma_1-1}\Big)^{1/(2-q_1)}
x^{1-2/[n(2-q_1)]}\ .
\end{eqnarray}
Since when $x\rightarrow+\infty$, $v_1(x)$ is either zero or
$v_1\rightarrow H_1x^{1-1/n}$ and $1-2/[n(2-q_1)]\neq1$ for any
values of $n$ and $q_1$, our asymptotic solutions will not lie on
the singular surface for fluid $1$. For ${\cal D}_2(x)=0$, we have
the similar result
\begin{eqnarray}
&v_2=\kappa nx-\Big(\kappa^{2q_2}\gamma_2\alpha_2^{q_2+\gamma_2-1}
x^{2q_2}\Big)^{1/(2-q_2)}\qquad\qquad\quad\nonumber\\
&\rightarrow \kappa nx-\Big(\gamma_2\kappa^{2q_2}
E_2^{q_2+\gamma_2-1}\Big)^{1/(2-q_2)}x^{1-2/[n(2-q_2)]}\ .
\end{eqnarray}
Since when $x\rightarrow+\infty$, $v_2$ is either zero or
$v_2\rightarrow H_2x^{1-1/n}$ and $1-2/[n(2-q_2)]\neq1$ for any
values of $n$ and $q_2$, our asymptotic solutions will not lie on
the singular surface for fluid $2$. In summary, for either fluid in
the regime of $x\rightarrow+\infty$, our asymptotic solutions will
not encounter the respective singular surfaces.

\subsection{Jump Conditions for Self-Similar Shocks}
\label{self-similar shock}

Large-scale shocks have been revealed in clusters of galaxies
through high-resolution X-ray imaging observations and radio
observations (e.g., Nulsen et al. 2005a, 2005b; McNamara et al.
2005; Bagchi et al. 2006). In order to probe and model such shock
features in clusters of galaxies, we construct numerically
semi-complete flow solutions with shocks across singular surfaces
in both hot gas and dark matter halo separately. These large-scale
shocks travel outward in a self-similar manner with asymptotic
flow signatures at large $r$ for given time $t$.

In principle, the self-similar form of shock radial positions
$r_{s,i}$ in fluid $i$ can be expressed either by the downstream
parameters ($K_i$ and $x_i$ on the downstream side of a shock) or by
the upstream parameters ($K_i$ and $x_i$ on the upstream side of a
shock). For simplicity, we express shock radial positions $r_{s,i}$
in terms of the upstream parameters. For fluid $i$, the shock radial
position is denoted by
\begin{eqnarray}
r_{s,i}=K_{u,i}^{1/2}t^nx_{u,i}\ ,
\end{eqnarray}
where $K_{u,i}$ is the upstream value of $K_i$ and $x_{u,i}$ is the
shock position in terms of the upstream independent self-similar
variable for fluid $i$. We introduce the sound speed ratio
$\lambda_i$ across a shock for fluid $i$ as
\begin{eqnarray}
\lambda_i=\bigg(\frac{K_{d,i}}{K_{u,i}}\bigg)^{1/2},
\end{eqnarray}
where the subscript $d$ denotes the downstream side (i.e., from the
shock position towards the centre) while the subscript $u$ denotes
the upstream side (i.e., from the shock position towards infinity).
In reference to a shock, $K_{d,i}$ is the parameter of $K_i$ on the
downstream side; $x_{d,i}$ is the shock location in terms of the
downstream self-similar variable for fluid $i$; $\rho_{d,i}$ is the
mass density downstream of a shock in fluid $i$ while $\alpha_{d,i}$
is the reduced mass density of a shock downstream of fluid $i$;
$P_{d,i}$ is the pressure downstream of a shock for fluid $i$ while
$\beta_{d,i}$ is the reduced pressure downstream of a shock for
fluid $i$; $u_{d,i}$ is the radial flow speed downstream of a shock
for fluid $i$ while $v_{d,i}$ is the reduced radial flow speed
downstream of a shock for fluid $i$; $M_{d,i}$ is the enclosed mass
downstream of a shock for fluid $i$ while $m_{d,i}$ is the reduced
enclosed mass downstream of a shock for fluid $i$. In one-to-one
correspondence, all these variables with subscript $\{u,\ i\}$ refer
to variables on the upstream side of a shock for fluid $i$.
In order to construct self-similar shocks, we require index
parameter $n$ to be the same across a shock to avoid unphysical
interface separation. As the shock radius is
$r_{s,i}=K_{u,i}t^nx_{u,i}$, the dimensional shock speed is then
given by
\begin{eqnarray}
u_{s,i}=\frac{\partial{r_{s,i}}}{\partial{t}}=nK^{1/2}_{u,i}t^{n-1}x_{u,i}
=n\frac{r_{s,i}}{t}\ .  \label{vshock}
\end{eqnarray}
Here, $u_{s,i}$ is the travel speed of a self-similar shock in fluid
$i$, indicating that the shock actually travels with a variable
speed for $n\neq 1$. For $n>1$ and $n<1$, the shock travel speed
increases (acceleration) and decreases (deceleration) with time $t$,
respectively. For $n=1$ and fairly arbitrary $\gamma_i$, such a
self-similar shock (not necessarily isothermal though) travels with
a constant speed. In clusters of galaxies, positions of shocks
observed vary from tens of kpcs (e.g., McNamara et al. 2005; Nulsen
et al. 2005b) to a few Mpcs (e.g., A3667 in Rotteringer et al. 1997;
galaxy cluster Abell 3376 in Bagchi et al. 2006; ZwCl 2341.1+0000 in
Bagchi et al. 2002; A786, A2255, A2256 in Ensslin et al. 1998).
In our scenario, such shocks actually travel to current positions
from a inner region around the cluster centre after their emergence,
which can be estimated in our model framework.
Then in our scenario at time $t$, the shock position $r_{s,i}$ is
determined by the following equation
\begin{eqnarray}
r_{s,i}=C_{s,i}\ t^n\ ,\label{shockposition}
\end{eqnarray}
where $C_{s,i}\equiv K_{u,i}^{1/2}x_{u,i}$ is a constant to be
estimated from observations at a certain time under specific
situations.

The self-similar transformation for variables on the downstream
side of a shock is
\begin{eqnarray}
x_{d,i}=\frac{r_{s,i}}{\lambda_iK_{u,i}^{1/2}t^n}\ ,
\end{eqnarray}
\begin{eqnarray}
\rho_{d,i}=\frac{\alpha_{d,i}(x_{d,i})}{4\pi Gt^2}\ ,
\end{eqnarray}
\begin{eqnarray}
P_{d,i}=\lambda_i^2\frac{K_{u,i}t^{2n-4}}{4\pi
G}\beta_{d,i}(x_{d,i})\ ,
\end{eqnarray}
\begin{eqnarray}
u_{d,i}=\lambda_iK_{u,i}^{1/2}t^{n-1}v_{d,i}(x_{d,i})\ ,
\end{eqnarray}
\begin{eqnarray}
M_{d,i}=\lambda_i^3\frac{K_{u,i}^{3/2}t^{3n-2}}
{(3n-2)G}m_{d,i}(x_{d,i})\ .
\end{eqnarray}
In the shock reference framework, the mass conservation equation
across a shock front is
\begin{eqnarray}
\rho_{d,i}(u_{s,i}-u_{d,i})-\rho_{u,i}(u_{s,i}-u_{u,i})=0\ ,
\end{eqnarray}
which can be rewritten conveniently as
\begin{eqnarray}
[\rho_i(u_{s,i}-u_i)]_u^d=0\ ;  \label{shockmass}
\end{eqnarray}
likewise, the radial momentum conservation then gives
\begin{eqnarray}
[P_i+\rho_i(u_{s,i}-u_i)^2]_u^d=0\ ;  \label{shockmomentum}
\end{eqnarray}
and the energy conservation equation leads to
\begin{eqnarray}
\bigg[\frac{\rho_i}{2}(u_{s,i}-u_i)^3
+\frac{\gamma_i}{(\gamma_i-1)}P_i(u_{s,i}-u_i)\bigg]_u^d=0\ ,
  \label{shockenergy}
\end{eqnarray}
where the pair of brackets denotes the difference of the argument on
the downstream (supercript $d$) and upstream (subscript $u$) sides
of a shock. As for a single fluid, we now introduce two new
variables $\Gamma_{d,i}$ and $\Gamma_{u,i}$ below
\begin{eqnarray}
\Gamma_{d,i}\equiv n-{v_{d,i}}/{x_{d,i}}\ ,
\end{eqnarray}
\begin{eqnarray}
\Gamma_{u,i}\equiv n-{v_{u,i}}/{x_{u,i}}\ .
\end{eqnarray}
Once we know the values of ($\Gamma$, $\alpha$, $x$) on the
downstream side (indicated by a subscript $d$) of a shock, we can
immediately calculate the corresponding variables on the upstream
side (indicated by a subscript $u$) or vice versa.
Details of shock calculations can be found in Appendix F. We only
show the major results here. The variable $\Gamma_{u,i}$ on the
upstream side can be computed from the variables on the downstream
side from the following equation
\begin{eqnarray}
\Gamma_{u,i}=\frac{2\gamma_i}{(\gamma_i+1)}
\alpha_{d,i}^{q_i+\gamma_i-1}\Gamma_{d,i}^{q_i-1}
x_{d,i}^{3q_i-2}+\frac{(\gamma_i-1)}{(\gamma_i+1)}\Gamma_{d,i}\ ;
\end{eqnarray}
other variables can be readily determined in a straightforward
manner
\begin{eqnarray}
\alpha_{u,i}=\frac{\alpha_{d,i}\Gamma_{d,i}}{\Gamma_{u,i}}\ ,
\end{eqnarray}
\begin{eqnarray}
\lefteqn{x_{u,i}=\Big(\alpha_{d,i}^{q_i+\gamma_i}
\Gamma_{d,i}^{q_i}x_{d,i}^{3q_i-2}+\alpha_{d,i}\Gamma_{d,i}^2
}\nonumber\\
&&{} -\alpha_{d,i}\Gamma_{u,i}\Gamma_{d,i}\Big)^{1/(3q_i-2)}
\bigg(\frac{\Gamma_{u,i}^{\gamma_i}} {\alpha_{d,i}^{q_i+\gamma_i}
\Gamma_{d,i}^{q_i+\gamma_i}}\bigg)^{1/(3q_i-2)}\ .
\end{eqnarray}
It then follows that
\begin{eqnarray}
v_{u,i}=x_{u,i}(n-\Gamma_{u,i})\
\end{eqnarray}
and the ratio $\lambda_i$ across a shock in fluid $i$ can be
determined accordingly. The upstream Mach number $\mathcal{M}_{u,i}$
is defined by
\begin{eqnarray}
\mathcal{M}_{u,i}&=&\frac{(u_{u,i}-u_{s,i})}{a_{u,i}}
=\bigg(\frac{\rho_{u,i}}{\gamma_iP_{u,i}}\bigg)^{1/2}
(u_{u,i}-u_{s,i})\nonumber\\
&=&\frac{-\Gamma_{u,i}}{\gamma_i^{1/2}
\alpha_{u,i}^{(q_i+\gamma_i-1)/2}x_{u,i}^{q_i-1}
(nx_{u,i}-v_{u,i})^{q_i/2}}\ ,
\end{eqnarray}
where $a_{u,i}$ is the polytropic sound speed on the upstream side
of a shock in fluid $i$, namely
\begin{eqnarray}
a_{u,i}=\bigg(\frac{\partial{P_{u,i}}}
{\partial{\rho_{u,i}}}\bigg)_s^{1/2} =\bigg(\frac{\gamma_i
P_{u,i}}{\rho_{u,i}}\bigg)^{1/2}\ .\label{soundspeed}
\end{eqnarray}
In general, shock positions in the two fluids are different and the
ratio $\kappa$ will change once either fluid goes across the
singular surface via a shock. Right at the shock front, we have
$r_{d,i}=r_{u,i}=r_{s,i}$,
$\lambda_i=(K_{d,i}/K_{u,i})^{1/2}=x_{u,i}/x_{d,i}$; we can then
recalculate $\kappa$
with the definition $\kappa\equiv (K_1/K_2)^{1/2}=x_2/x_1$ where
$K_i$ should take the local value.

The specific `entropy' of fluid $i$ is conserved along streamlines
and is given by
\begin{eqnarray}
s_i&=&c_{v,i}\ln\bigg(\frac{P_i}{\rho_i^{\gamma_i}}\bigg)\nonumber\\
&=&c_{v,i}\ln\bigg\{K_i^{(4-3\gamma_i)/(3n-2)}\\ \nonumber
&&\qquad\qquad\quad\times(4\pi
G)^{\gamma_i-1}[(3n-2)G]^{q_i}M_i^{q_i}\bigg\}\ ,
\end{eqnarray}
where $c_{v,i}$ is the specific heat capacity at constant volume for
fluid $i$. As the specific entropy increases from upstream side to
downstream side across a shock, either parameter $K_i$ increases
from upstream side to downstream side for $\gamma_i<4/3$, or
parameter $K_i$ decreases from upstream side to downstream side for
$\gamma_i>4/3$; we take $\gamma_i\neq 4/3$ in this paper. For a
single self-gravitating polytropic gas with $\gamma=4/3$, the reader
is referred to Lou \& Cao (2007) for a further theoretical
development of earlier analyses by Goldreich \& Weber (1980) and
Yahil (1983).

\section{Results of Numerical Examples}

Up to now, within the self-similar dynamic model framework of two
gravity coupled polytropic fluids, we have successfully constructed
the generalized version of quasi-static asymptotic solutions for
small $x$ in reference to the model analysis of Lou \& Wang (2006,
2007). This type of self-similar evolution eventually approaches a
static configuration with a diverging density towards to the central
core region. Meanwhile, we have determined the two singular surfaces
and the shock conditions across the two singular surfaces,
respectively. In order to construct a global semi-complete
quasi-static solution, the relevant parameters required to be known
are: the scaling index $n$ as introduced in self-similar
transformation (\ref{transformation}); a proper starting value
$x_{ini}$ in the small $x$ regime to guarantee a reliable numerical
integration; the two polytropic indices of dark matter $\gamma_1$
and of hot ICM $\gamma_2$ respectively; an estimate of time $t$ when
a self-similar evolution is presumed to begin; the two parameters
$K_1$ and $K_2$ are related to the sound speeds of two fluids
respectively and are introduced in self-similar transformation
(\ref{transformation}) or equivalently, the parameter $K_2$ and the
ratio $\kappa\equiv (K_1/K_2)^{1/2}$ of the two fluids; the initial
parameters for the quasi-static velocities $L_1$ and $L_2$ as
defined in the quasi-static solution (\ref{quasiv}); the independent
self-similar variables on the downstream sides of shock positions
for dark matter $x_{d,1}$ and for hot ICM $x_{d,2}$. Once these
eleven parameters are specified, a semi-complete numerical solution
can be established. If we just construct a dimensionless solution,
then values of $K_2$ and $t$ are not needed, indicating that only
nine dimensionless parameters are required to be known. Note that
parameter $x_{ini}$ needs to be carefully chosen. Flow parameters at
large $x$ can be determined accordingly.

Now we try to use our two-fluid model to explore dynamic behaviours
of hot ICM and dark matter halo for clusters of galaxies. Throughout
this paper, we take fluid $1$ to represent the dark matter halo and
fluid $2$ to represent the hot fully ionized ICM. Any variables with
subscript $2$ are associated with the hot ICM in galaxy clusters in
the model analysis. We take the quasi-static solution of both fluids
as $x\rightarrow0^+$ and go across the singular surfaces with shocks
in hot gas and in dark matter halo (at different locations and thus
different outward shock travel speeds) respectively. The
cluster-scale shocks have been observed in many clusters of
galaxies, which may be related to cluster formation processes,
central AGN activities or merging of galaxies. Although shocks in a
dark matter halo have not yet been detected, there is no obvious
reason to rule out this possibility. Observationally, it may be
possible to test their presence once the density jump profile of a
dark matter halo can be inferred through effects of gravitational
lensing. Very recently, Onemli \& Sikivie (2007) proposed to
interpret certain gravitational lensing observations to be
``caustics" (i.e., sharp rises of density in DM halos) in galaxy
clusters. While dark matter shocks in our model differ from such DM
caustics discussed in the literature, they do share certain similar
features and therefore, shocks may also be detected by utilizing
gravitational lensing effects. More detailed discussion on DM
caustics can be found in the Discussion section at the end. In our
model, we can describe various dynamical behaviours of hot ICM when
$r\rightarrow +\infty$, including inflow, outflow and static
solutions. Especially for the outflow solutions at large $x$, we
shall refer to them as {\it galaxy cluster winds}, just like solar
and stellar winds or galactic winds on much smaller yet different
scales. On the basis of galaxy cluster winds and flows of dark
matter halo, we will estimate the loss of baryon matter during a
timescale of the order of $\sim 10^9$ yr for the evolution of galaxy
clusters.

As scaling index $n$ is a key parameter controlling asymptotic
scaling behaviours of self-similar dynamic solutions, we shall
discuss model solutions for three cases of $2/3<n<1$, $n=1$ and
$n>1$, respectively, all with $\gamma_i$ in the range of
$1<\gamma_i<2$ (polytropic indices $\gamma_1$ and $\gamma_2$ of
two fluids are allowed to be different in general). We use the
standard fourth-order Runge-Kutta numerical scheme (e.g., Press et
al. 1986) to integrate coupled nonlinear ODEs (\ref{a1}),
(\ref{dv1}), (\ref{da2}) and (\ref{v2}) with relevant asymptotic
solutions in both regimes of large and small $x$. As the analytic
quasi-static solution is adopted for $x\rightarrow 0^{+}$ in this
model consideration, we start the numerical integration with an
assigned value of $x_{ini}$, which is small enough.\footnote{The
rule of thumb criterion for a sensible choice of $x_{ini}$ is that
$v_i(x)/x^k$ remains constant for a certain range of small $x$.}
The corresponding initial values of $\alpha_i(x)$ and $v_i(x)$ are
determined by
quasi-static solution (\ref{quasia}) and (\ref{quasiv}) with
mutually consistent coefficients. Once a fluid component
encounters its singular surface, we let the fluid go across it via
a shock; there is a certain degree of freedom in choosing shock
location and thus shock speed in constructing solutions. If the
solution can be integrated to infinity (a sufficiently large $x$
in practice), a semi-complete solution is then obtained in our
two-fluid model. Note that in order to avoid unstable numerical
integration from small $x$ outwards, we need to take some
appropriate initial parameters and carefully check characteristic
features of a true quasi-static solution at small $x$.

In applications to clusters of galaxies, the virial radius
corresponds to the radius where the mass density of dark matter is
200 times of the critical density $\rho_c$ in the Universe (e.g.,
Navarro et al. 1996). We follow this convention and denote the
virial radius by $r_{200}$. Here, the critical mass density
$\rho_{c}$ is defined by
\begin{eqnarray}
\rho_c={3H_0^2}/(8\pi G)\ ,
\end{eqnarray}
where $H_0$ is the Hubble constant and $G=6.67\times10^{-11}\
\hbox{kg}^{-1}\hbox{m}^3\hbox{ s}^{-2}$ is the gravitational
constant. Recent {\it Wilkinson Microwave Anisotropy Probe (WMAP)}
results showed a Hubble constant $H_0=72\pm 8$ km s$^{-1}$
Mpc$^{-1}$ (e.g., Freedman et al. 2001; Spergel et al. 2003). Thus
the critical mass density is currently inferred to be
$\rho_c=9.7\times10^{-27}$ kg m$^{-3}$. In our discussion on
clusters of galaxies, the radius $r$ will be expressed in unit of
the virial radius $r_{200}$.

\subsection{Ranges of $K_2$ and $t$ Values}

Once values\footnote{As an example, we take $K_2\equiv K_{d,2}$ in
this subsection for the hot ICM on the downstream side of an ICM
shock.}
of $K_2$ and time $t$ are chosen in our two-fluid model, we can
then simulate a dynamic shock flow of hot ICM in clusters of
galaxies under the gravitational influence of a flowing dark
matter halo. Physically, we may take $t$ to be the timescale when
a large-scale shock has emerged. As shocks in clusters of galaxies
may be initiated by gravitational core collapses or AGN activities
or merging galaxies and the timescale for the recurrence of AGNs
is of the order of $\sim 10^8$ yr (or equivalently $\sim 2\times
10^{15}$ to $2\times 10^{16}$ s, e.g., Fabian 1994), we may choose
time $t$ of this magnitude order in our model applications to
clusters of galaxies. According to dimensionless equation of state
(\ref{stateless}) and self-similar transformation
(\ref{transformation}), we have
\begin{eqnarray}
K_2=\bigg[\frac{\rho_2^{\gamma_2}}{P_2} M_2^{q_2}(4\pi)^{\gamma_2-1}
G^{q_2+\gamma_2-1}(3n-2)^{q_2}\bigg]^{2/(3q_2-2)}\ . \label{eq96}
\end{eqnarray}
For a hot ICM, we simply apply the ideal gas law $P_2={\cal
N}_2k_BT_2$,
where ${\cal N}_2$ is the particle number density, $k_B$ is the
Boltzmann constant, $P_2$ is the thermal pressure of ICM, and $T_2$
is the ICM temperature. Then equation (\ref{eq96}) appears as
\begin{eqnarray}
K_2=\bigg[\frac{\rho_2^{\gamma_2}}{{\cal N}_2k_BT_2}M_2^{q_2}
(4\pi)^{\gamma_2-1}\qquad\qquad\qquad\qquad\nonumber\\
\qquad\qquad\qquad\qquad \times
G^{q_2+\gamma_2-1}(3n-2)^{q_2}\bigg]^{2/(3q_2-2)}\ . \label{K_22}
\end{eqnarray}
As in clusters of galaxies the electron number density is typically
$10^{-2}\sim 10^{-4}\hbox{ cm}^{-3}$ (e.g., Fabian 1994; Cavaliere
\& Fusco-Femiano 1978; Nulsen et al. 2005b) and the gas mainly
consists of protons, electrons, and $\alpha$ particles (nuclei of
helium atoms), the particle number density is also of this
magnitude. The mean molecular weight for ICM in galaxy clusters is
about $0.6\hbox{ g/mol}$ (e.g., Cavaliere \& Fusco-Femiano 1978) and
the mass of hot ICM in galaxy clusters is of the order of
$10^{13}M_{\odot}$ (e.g., Peres et al. 1998). Typically, the ICM
temperature $T_2$ varies in the range of $\sim 10^7-10^8$ K (e.g.,
Fabian 1994). For an observational input of these different
parameters, we can then estimate the typical range of $K_2$ values
(in SI unit). To be specific, we estimate relevant parameters for
our two-fluid model calculations and the results are summarized in
Table \ref{K_2} above.
\begin{table}
\begin{center}
\caption{Values of $K_2$ in SI unit
for different sets of model parameters. We take the mean molecular
weight of the typical ICM in galaxy clusters to be $0.59\hbox{
g/mol}$, corresponding to the total mass ratio of protons to
$\alpha$ particles being 3.
Then the ICM mass density $\rho_2$ can be calculated from $N_e$,
the electron number density in the hot ICM, and $K_2$ is
calculated from equation (\ref{K_22}). The temperature $T_2$ is in
unit of keV and the electron number density $N_e$ is in unit of
cm$^{-3}$. The enclosed mass $M_2$ for ICM is in unit of
$10^{13}M_{\odot}$ ($M_{\odot}$ is the solar mass). All values of
$K_2$ in Table 1 are in SI unit.
}
 \label{K_2}
\begin{tabular}{|c|c|c|c|c|c|}
  \hline
  $n$ & $\gamma_2$ & $T_2$ & $N_e$ & $M_2$ & $K_2$ \\
  \hline\hline
  0.8 & 1.31 & 6 & 0.004 & 1 & $1.11\times10^{21}$ \\
  0.8 & 1.31 & 7 & 0.004 & 1 & $2.68\times10^{21}$ \\
  0.8 & 1.31 & 7 & 0.008 & 1 & $7.84\times10^{20}$ \\
  \hline
  1 & 1.405 & 8 & 0.002 & 1 & $3.72\times10^8$ \\
  1 & 1.405 & 7 & 0.002 & 1 & $6.93\times10^8$ \\
  1 & 1.405 & 8 & 0.001 & 2 & $1.37\times10^9$ \\
  \hline
  1.07 & 1.42 & 5 & 0.001 & 1 & $9.29\times10^6$ \\
  1.07 & 1.42 & 6 & 0.001 & 1 & $3.98\times10^6$ \\
  1.07 & 1.42 & 6 & 0.003 & 1 & $3.41\times10^7$ \\
  \hline
\end{tabular}
\end{center}
\end{table}

\subsection{The ICM Temperature Profile}

According to the ideal gas law $P_2=\rho_2 k_BT_2/\mu_0$, where
$\mu_0$ is the mean molecular mass, $k_B$ is the Boltzmann
constant and $\rho_2$ is the ICM mass density, the ICM temperature
$T_2$ in our self-similar flow model is given by
\begin{eqnarray}
T_2(r,t)=\frac{\mu_0K_2}{k_B}\ t^{2n-2}\kappa^{2q_2}
\alpha_2^{q_2+\gamma_2-1}x^{2q_2}(\kappa nx-v_2)^{q_2}\ . \label{T}
\end{eqnarray}
For $x\rightarrow0^{+}$, the asymptotic behaviour of ICM temperature
is then
\begin{eqnarray}
T_2\rightarrow \frac{\mu_0K_2}{k_B}\ t^{2n-2}\kappa^{3q_2}
A_2^{q_2+\gamma_2-1}n^{q_2}x^{2-2/n}+\ldots\ , \label{Tsmall}
\end{eqnarray}
which diverges as $r\rightarrow 0^+$ for $n<1$. As the radial flow
speed $v_i(x)$ is small compared with $x$ as $x\rightarrow+\infty$
(see equations (\ref{largev1}), (\ref{largev2}), (\ref{largev12}),
(\ref{largev22})) and the reduced mass density $\alpha_i(x)$ has
similar asymptotic scaling behaviours at both large and small $x$
(see equations (\ref{largea1}), (\ref{largea2}) and
(\ref{quasia})), the ICM temperature $T_2$ has the similar
asymptotic scaling behaviours at both large and small $x$ because
$T_2$ is closely related to thermodynamic variables $P_2$ and
$\rho_2$ as well as the enclosed mass $M_2$. Therefore for $n<1$,
the ICM temperature $T_2$ decreases with increasing radius $r$; in
the limit of $n\rightarrow (2/3)^{+}$, the limiting temperature
scaling would be $\sim r^{-1}$. In fact, qualitatively similar
temperature profiles have been observed in several galaxy clusters
(e.g., galaxy clusters A2256, A2319, A665 in Markevitch 1996; and
galaxy cluster A520 in Markevitch et al. 2005).
Both temperature and mass density diverge as $x\rightarrow 0^{+}$
for $n<1$; we need to introduce a sensible reference radius to
cutoff around the cluster centre.

With $n=1$, expression (\ref{Tsmall}) for ICM temperature $T_2$
remains constant for a certain radial distance around the centre.
This kind of solution actually represents galaxy clusters with
roughly constant temperature. This kind of galaxy clusters has
also in fact been observed, such as galaxy cluster A2204 in Peres
et al. (1998) and galaxy clusters A2199 and 2A 0335+096 in Sanders
\& Fabian (2006). For the $n=1$ case of our two-fluid model,
solutions are not `isothermal' in general because $\gamma_i\neq
1$; and this clearly differs from previous isothermal analyses
with $n=1$ and $\gamma=1$ (e.g., Tsai \& Hsu 1995; Lou \& Shen
2004; Lou 2005; Bian \& Lou 2005). Of course, we could set
$\gamma_1=1$ and $\gamma_2\neq 1$ for an `isothermal' dark matter
flow and a nonisothermal gas or vice versa or
$\gamma_1=\gamma_2=1$ for two `isothermal' flows and so forth. In
particular, we emphasize that this $n=1$ case fundamentally
differs from the condition $n=1$ in Suto \& Silk (1988), because
their specific entropy is not conserved along streamlines with
$n=1$.

For $n>1$, the ICM temperature $T_2$ increases with increasing
radius as indicated by expressions (\ref{T}) and (\ref{Tsmall}).
Temperature profiles in many clusters of galaxies do appear to
behave in this manner qualitatively in a certain radial range, such
as galaxy clusters Ms$0735.6+7421$ (e.g., McNamara et al. 2005),
Perseus (e.g., Sanders \& Fabian 2007) and A2052 (e.g., Blanton et
al. 2001).

\subsection{Energetics of the Coupled Two-Fluid System}

Parallel to the single flow system of an isothermal gas (e.g., Tsai
\& Hsu 1995), the energy of our coupled polytropic two-fluid system
consists of three parts, namely, the gravitational potential energy
denoted by $E_{grav}$, the kinetic energy of two fluids denoted by
$E_{k,i}$ and the thermal energy of two fluids denoted $E_{th,i}$,
where the subscript $i=1,\ 2$ refer to fluid 1 (dark matter) and
fluid 2 (hot ICM) respectively. In conventional scenarios, AGN
activities are sustained by accretions of baryon matter onto a
supermassive black hole (SMBH). Conceptually, it is also physically
sensible to think of accretions of both baryon matter and dark
matter onto a SMBH for AGN activities (e.g., Hu et al. 2006),
although only radiations from ICM can directly reach us. Therefore
in addition to radiative losses from electrons we observe, the
outburst energy of an AGN should also involve the energies
associated with hot gas and dark matter. The energy within the shock
radius contains the outburst energy of AGN and the original energy
in the system prior to an AGN. We may calculate the energy within
the shock radius to estimate the order of the outburst energy
released by an AGN.

The gravitational energy of our coupled two-fluid system within a
radial range between $r_{min}$ and $r_{max}$ (the respective
dimensionless self-similar variables are $x_{min}$ and $x_{max}$ for
a specified time $t$) is simply given by
\begin{eqnarray}
E_{grav}=-\int_{r_{min}}^{r_{max}}
\frac{G(M_1+M_2)}{r}(\rho_1+\rho_2) 4\pi r^2dr\ .
\end{eqnarray}
With self-similar transformation (\ref{transformation}), we then
derive
\begin{eqnarray}
&&\!\!\!\!\!\! E_{grav}=-\frac{\kappa^5K_2^{5/2}t^{5n-4}}
{(3n-2)G}\times\qquad\qquad\qquad\qquad\nonumber\\
&&\!\!\!\!\!\!\!\!\!\!\!\!\!
\int_{x_{min}}^{x_{max}}x^3\bigg[\alpha_1(nx-v_1)
+\alpha_2\bigg(nx-\frac{v_2}{\kappa}\bigg)\bigg]
(\alpha_1+\alpha_2)dx\ , \label{genergy}
\end{eqnarray}
where $x\equiv x_1$ for dark matter halo. The kinetic energy for
fluid $i$ within the radial range between $r_{min,i}$ and
$r_{max,i}$ (the respective dimensionless self-similar variables are
$x_{min,i}$ and $x_{max,i}$ for a specified time $t$) is simply
given by
\begin{eqnarray}
E_{k,i}&=&\int_{r_{min,i}}^{r_{max,i}}\frac{\rho_iu_i^2}{2}
\ 4\pi r^2dr\nonumber\\
&=&\frac{K_i^{5/2}t^{5n-4}}{2G}\int_{x_{min,i}}^{x_{max,i}}
\alpha_iv_i^2x_i^2dx_i\
.\label{kenergy}
\end{eqnarray}
The thermal energy of fluid $i$ within the same radial range of
$r_{min,i}$ and $r_{max,i}$
is simply given by
\begin{eqnarray}
E_{th,i}&=&\int_{r_{min,i}}^{r_{max,i}}
\frac{\gamma_iP_i}{(\gamma_i-1)}\ 4\pi r^2dr\nonumber\\
&=& \frac{\gamma_i K_i^{5/2}t^{5n-4}}{(\gamma_i-1)G}
\int_{x_{min,i}}^{x_{max,i}}x_i^2\alpha_i^{\gamma_i}m_i^{q_i}dx_i\
.\label{thenergy}
\end{eqnarray}
As the polytropic sound speed $a_i$ in fluid $i$ is
\begin{eqnarray}
a_i=\Big(\gamma_i\frac{P_i}{\rho_i}\Big)^{1/2}
=\Big(\gamma_i\frac{\beta_i}{\alpha_i}\Big)^{1/2}
K_i^{1/2}t^{n-1}\
,
\end{eqnarray}
the common dimensional coefficient in integrals
(\ref{genergy})$-$(\ref{thenergy}) can be expressed in terms of the
sound speed as
\begin{eqnarray}
\frac{K_i^{5/2}t^{5n-4}}{G}=\frac{a_i^5t}{G}\
\Big(\gamma_i\frac{\beta_i}{\alpha_i}\Big)^{-5/2}\ .
\end{eqnarray}
In our model, the sound speed $a_i$ is not a constant, which differs
from the isothermal model of Tsai \& Hsu (1995) for a single gas.

The total energy of the two-fluid system within the radial range
between $r_{min}$ and $r_{max}$ is then
\begin{eqnarray}
E_{total}=E_{grav}+\sum_{i=1}^2E_{k,i}+\sum_{i=1}^2E_{th,i}\
.\label{energy}
\end{eqnarray}

At a certain reference time $t_1$, we can calculate the total
energy $E_{total, o}$ within a radial range between $r_{min}$ and
$r_{max}$, where neither shock arrives. After a certain time lapse
$t_2$ when both shocks have passed through the radial range under
consideration, we can calculate the total energy $E_{total,f}$
within the same radial domain. Then the energy difference
$E_{total,f}-E_{total,o}$ is the energy input from shock flow.
Furthermore, the mean input power during this time interval
$t_2-t_1$ can be estimated by
\begin{eqnarray}
{\cal P}_{total}=(E_{total,f}-E_{total,o})/(t_2-t_1)\ .
\label{power}
\end{eqnarray}
In addition to radiation losses from the central region, this mean
input power can be used to estimate the power from AGN activities.
In our model framework, the total energy includes several parts. The
X-ray luminosity of galaxy clusters inferred by observations (e.g.,
Nulsen et al. 2005a; Gizani \& Leahy 2004) should come from the
thermal energy part of hot ICM. In the estimated outburst energies
of AGNs with certain models (e.g, Nulsen et al. 2005a; Gizani \&
Leahy 2004), only the fraction of energy transmitted to the thermal
reservoir of hot ICM is considered. In our scenario, the part of
released energy from AGNs is also transmitted to the dark matter as
well as the gravitational and kinetic energies of hot ICM.
Therefore, our mean power and energy are more than just the thermal
energy input to hot ICM.

\subsection{Solution Examples of $2/3<n<1$}

In this case of $2/3<n<1$, features of our global semi-complete
solutions are summarized as follows. When $x\rightarrow +\infty$,
$v_i(x)$ goes to zero with a $\alpha_i(x)$ scaling of $x^{-2/n}$.
In the other limit of $x\rightarrow0^{+}$, $v_i(x)$ goes to zero
with a $\alpha_i(x)$ scaling of $x^{-2/n}$. At large $x$, both
fluids may have various combinations of asymptotic behaviours of
outflow, breeze, static envelope, contraction, inflow as the two
shock positions (or outgoing speeds) vary. Through numerical
exploration, we note that as the shock position decreases from the
vicinity of the singular surface to smaller $x$ in fluid $i$, the
corresponding outflow speed $v_i(x)$ at large $x$ tends to
decrease. If the shock position is reduced further, the fluid at
large $x$ may become static and then turn into a contraction or an
inflow. As implied by the limit of our parameter regime for
scaling index $n$, the most steep density scaling both at large
and small $x$ would be $\sim x^{-3}$ (Wang \& Lou 2007). This
limiting density scaling can be systematically tested against
observations of galaxy clusters.

\begin{figure}
\includegraphics[height=6cm]{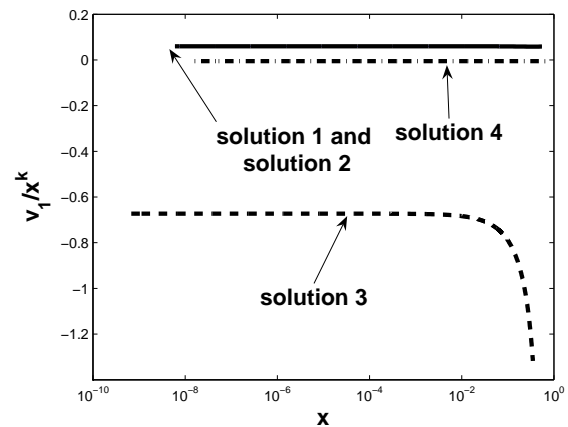}
\caption{\label{nsmall1_1} Scaled radial flow velocity of fluid $1$
(dark matter halo) at small $x$ with $2/3<n<1$.
The relevant model parameters are summarized in Table \ref{nsmall1}
for solutions 1, 2, 3 and 4. Note that for a certain range of
initial distance, $v_1(x)/x^k$ remains a constant, showing that
these solutions for $v_1(x)$ are indeed quasi-static for small $x$.
For all these solutions 1, 2, 3 and 4, we have the same $\{n,\
\gamma_1,\ \gamma_2,\ \kappa\}=\{0.8,\ 1.3,\ 1.31,\ 0.02\}$.
Solutions 1 and 2 (heavy solid line) coincide in the regime of small
$x$. The corresponding solution behaviours of $v_2(x)/x^k$ (for the
hot ICM) are shown in Fig. \ref{nsmall1_2}.
}
\end{figure}

\begin{figure}
\includegraphics[height=6cm]{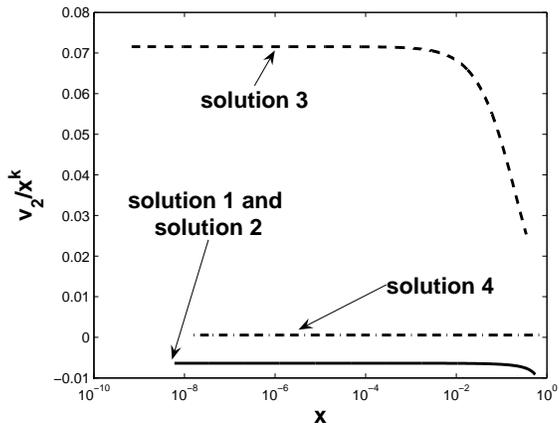}
\caption{\label{nsmall1_2} Scaled radial flow velocities of fluid
$2$ (hot ICM) at small $x$ with $2/3<n<1$; the relevant model
parameters are listed in Table \ref{nsmall1}. Note that for a
certain range of initial distance, $v_2(x)/x^k$ remains constant,
showing that these solutions 1, 2, 3 and 4 for $v_2(x)$ are indeed
quasi-static for small $x$. Together with Figure \ref{nsmall1_1} for
fluid 1 (dark matter halo), we have succeeded in constructing
solutions with both fluids being quasi-static at small $x$. For all
these solutions 1, 2, 3 and 4, we have the same $\{n,\ \gamma_1,\
\gamma_2,\ \kappa\}=\{0.8,\ 1.3,\ 1.31,\ 0.02\}$. Solutions 1 and 2
(heavy solid line) coincide in the regime of small $x$. }
\end{figure}

Here, we offer a few solution examples with the set of parameters
$\{n,\ \gamma_1,\ \gamma_2,\ \kappa\}=\{0.8,\ 1.3,\ 1.31,\
0.02\}$.
The two coefficients $A_1$ and $A_2$ can be readily calculated
from equation (\ref{staticdensity}) and the value of index $k$ is
determined by quartic equation (\ref{k}) (we choose the root $k>1$
for the consistency of series expansion analysis); their specific
values are $\{A_1,\ A_2,\ k\}=\{1.76,\ 0.35,\ 1.99\}$
correspondingly. For different initial values and shock positions,
the global semi-complete solutions for radial flow speeds $v_i(x)$
(scaled for a clear presentation) are shown in Figures
\ref{nsmall1_1} and \ref{nsmall1_2}; numerical values of the
relevant solution parameters are listed in Table \ref{nsmall1}.
In short, for these four numerical solutions 1, 2, 3 and 4, the nine
parameters to determine a dimensionless solution are: $\{n,\
\gamma_1,\ \gamma_2,\ \kappa,\ L_1,\ L_2,\ x_{ini},\ x_{d,1},\
x_{d,2}\}$=$\{0.8,\ 1.3,\ 1.31,$ $\ 0.02,\ 0.06,\ -0.00638,\
6\times10^{-9},\ 342,\ 0.55\}$ for solution 1; $\{n,\ \gamma_1,\
\gamma_2,\ \kappa,\ L_1,\ L_2,\ x_{ini},\ x_{d,1},\
x_{d,2}\}$=$\{0.8,\ 1.3,\ 1.31,$ $\ 0.02,\ 0.06,\ -0.00638,\
6\times10^{-9},\ 1.8,\ 0.63\}$ for solution 2; $\{n,\ \gamma_1,\
\gamma_2,\ \kappa,\ L_1,\ L_2,\ x_{ini},\ x_{d,1},\
x_{d,2}\}$=$\{0.8,\ 1.3,\ 1.31,$ $\ 0.02,\ -0.673,\ 0.0716,\
6.8\times10^{-10},\ 0.35,\ 50\}$ for solution 3; $\{n,\ \gamma_1,\
\gamma_2,\ \kappa,\ L_1,\ L_2,\ x_{ini},\ x_{d,1},\
x_{d,2}\}$=$\{0.8,\ 1.3,\ 1.31,$ $\ 0.02,\ -0.0051,\ 0.000542,\
1.6\times10^{-8},\ 0.8,\ 5\}$ for solution 4. It should be noted
that parameter $x_{ini}$ is not intrinsic to the physical
description but must be properly chosen within a certain range for
reliable numerical solutions matched with quasi-static asymptotic
solutions at small $x$.

\begin{table}
\begin{center}
\caption{Parameters for numerical solutions 1, 2, 3 and 4 presented
in Figures \ref{nsmall1_1} and \ref{nsmall1_2} with $2/3<n<1$. In
the first column on the left, `No.' is the numeral label to
distinguish different example solutions 1, 2, 3 and 4. `Type 1' is
the type of flow solutions of fluid $1$ when $x\rightarrow+\infty$,
and `Type 2' is the type of flow solutions of fluid $2$ when
$x\rightarrow+\infty$. Here, $x_{d,1}$ and $x_{d,2}$ are the
independent self-similar variables on the downstream sides of shock
positions for fluids $1$ and $2$, respectively.
Parameters $E_1$ and $E_2$ are respectively the coefficients of
$\alpha_1(x)$ and $\alpha_2(x)$ at a large enough $x$ according to
asymptotic solutions (\ref{largea1}) and (\ref{largea2}). Parameters
$H_1$ and $H_2$ are respectively the velocities $v_1(x)$ of fluid
$1$ and $v_2(x)$ of fluid $2$ at a large enough $x$ according to
asymptotic solutions (\ref{largev1}) and (\ref{largev2}). Parameter
$x_{ini}$ is the initial value of $x$ for a numerical integration.
As this table is too long horizontalwise, we break the table in two
parts and stack them together. }
\label{nsmall1}
\begin{tabular}{|c|c|c|c|c|c|c|c}
 \hline
 No.&Type 1&Type 2&$L_1$&$L_2$&$x_{ini}$\\
 \hline\hline
 1&outflow&outflow&0.06&$-0.00638$&$6\times10^{-9}$\\
 \hline
 2&inflow&outflow&0.06&$-0.00638$&$6\times10^{-9}$\\
 \hline
 3&outflow&inflow&$-0.673$&0.0716&$6.8\times10^{-10}$\\
 \hline
 4&inflow&inflow&$-0.0051$&0.000542&$1.6\times10^{-8}$\\
 \hline\hline
  $x_{d,1}$&$x_{d,2}$&$E_1$&$E_2$&$H_1$&$H_2$\\
 \hline\hline
  342&0.55&$6.04\times10^5$&0.432&$1471$&178\\
 \hline
  1.8&0.63&2.2&0.21&$-0.13$&0.034\\
 \hline
  0.35&50&6.22&877&48.5&$-7.61$\\
 \hline
  0.8&5&6.99&2.67&$-3.44$&$-0.0269$\\
 \hline
 \end{tabular}
 \end{center}
\end{table}

Solutions such as example solution $2$ catch our special attention.
Such solutions demonstrate a possible scenario that when
$x\rightarrow +\infty$, the hot ICM represented by fluid $2$ can
flow outward while the dark matter halo represented by fluid $1$
gradually contracts. As the dark matter halo dominates in clusters
of galaxies in terms of mass (the enclosed mass ratio of dark matter
halo to hot ICM varies from 4 to 10 in typical clusters of galaxies,
e.g., Peres et al. 1998), one can readily understand the behaviour
of this kind of dynamic solutions. Since the hot gas in this
solution can overcome the overall gravity (dark matter and hot ICM
together) and flow outward at large distance, we naturally refer to
such gas outflow as the {\it galaxy cluster wind}. The driving
energy of the galaxy cluster wind may come from the gravitational
core collapses or AGN activities or merging of galaxies around the
centre of galaxy clusters; a self-similar phase gradually emerges
after a while of dynamic evolution. For $x\rightarrow 0^+$, the flow
velocity of dark matter halo is outward. The outflow of dark matter
may be propelled by violent outbursts of energy around the cluster
centre occurred at earlier times.

In fact, hydrodynamic models for clusters of galaxies have been
proposed earlier. In Gunn \& Gott (1972), a theory of infall of
materials into clusters of galaxies was developed and they applied
it to the growth of galaxy clusters and the generation of hot ICM.
Bertschinger (1989) discussed the time-dependent evolution of
cooling flows in clusters of galaxies and emphasized that although
the size of a cooling flow region will increase with time, the gas
material itself does not go outward; this mechanism is referred to
as `cooling waves'. In particular, steady winds from a galaxy
clusters (e.g., the Coma cluster) have been discussed by Yahil \&
Ostriker (1973). Their result indicates a mass loss rate of $\sim
10^3-10^4M_{\odot}$ per year. Their model considered a steady-state
wind without involving shocks. In our dynamic model, we can
construct self-similar solutions such as solution 2 to explore the
dynamics of galaxy cluster winds with shocks and possible physical
consequences.

\begin{figure}
\includegraphics[height=6cm]{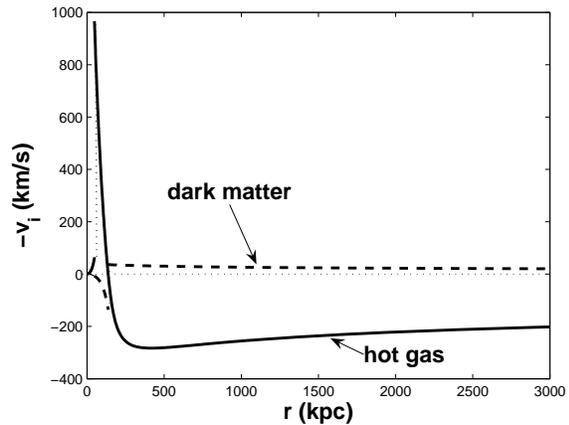}
\caption{\label{nsmall1_3_v}Negative radial flow velocities of hot
ICM (solid curve) and dark matter halo (dashed curve) of solution 2
with $t=7.59\times10^{15}$ s and $K_2=5.50\times10^{20}$ SI unit on
the downstream side of a shock in hot ICM. The shock position in the
hot ICM is at $\sim 48.38$ kpc while the shock position of dark
matter is at $\sim 138.23$ kpc. The initial parameters for the
numerical integration are listed in Table \ref{nsmall1} (see those
for solution 2). The corresponding radial flow velocities at small
$x$ of this solution are shown in Figs. \ref{nsmall1_1} and
\ref{nsmall1_2} (solution 2).}
\end{figure}
\begin{figure}
\includegraphics[height=6cm]{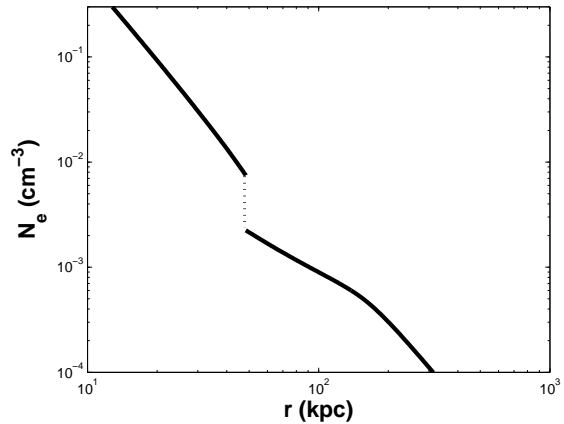}
\caption{\label{nsmall1_3_d} Electron number density profile for
$t=7.59\times10^{15}$ s and $K_2=5.50\times10^{20}$ SI unit on the
downstream side of a shock for solution 2 (see Figs. 1 and 2 for
details of the quasi-static solution behaviour at small $x$). The
corresponding global radial flow velocities of this solution are
shown in Fig. \ref{nsmall1_3_v}. Here we take the typical mean
molecular weight in clusters of galaxies to be $0.59$ g/mol. As the
electron number density can be obtained by high-resolution X-ray
imaging observations, we give the model electron number density so
that comparisons and tests can be made.}
\end{figure}

For quasi-static solution 2 in small $x$ shown in Figures 1 and 2,
we adopt estimates for the physical parameters $t=7.59\times10^{15}$
s and $K_2=5.50\times10^{20}$ SI unit on the downstream side of the
ICM shock. In this example, the shock position in the hot gas is
$r_{s,2}=48.38$ kpc (shock positions vary from tens of kpc to
hundreds of kpc in clusters of galaxies and a similar shock position
has been observed in the Perseus cluster; e.g., Fabian et al. 2006).
We only apply our model to radii less than several Mpcs, say $\sim
3$ Mpc (e.g., the size scale of the Hydra A cluster is $\sim 3$ Mpc;
see e.g., Taylor et al. 1990). The corresponding flow velocity
profiles are displayed in Figure \ref{nsmall1_3_v} and the profile
of electron number density $N_e$ is displayed in Figure
\ref{nsmall1_3_d}. In order to get the electron number density, we
take the mean molecular weight to be $\sim 0.59$ g/mol (see
Cavaliere \& Fusco-Femiano 1978). In this example, the radial
outflow velocity of hot gas at 3Mpc is $\sim 202$ km s$^{-1}$ while
the radial inflow velocity of dark matter halo at 3Mpc is $\sim
20.3$ km s$^{-1}$.
The analytical expression for the polytropic sound speed ratio can
be found in Appendix G and the sound speed ratio in this specific
example is displayed in Fig. \ref{nsmall1_soundratio}.

\begin{figure}
\includegraphics[height=7cm]{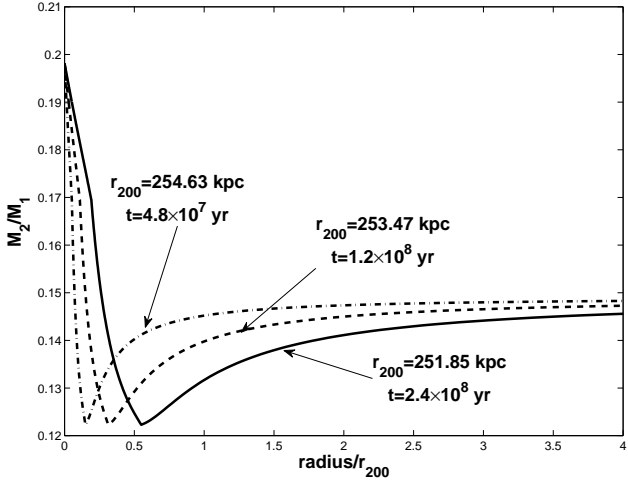}
\caption{\label{nsmall1_massratio}
The enclosed mass ratio $M_2/M_1$ as a function of $r$ at different
times of solution $2$. The abscissa is the radius in the unit of
virial radius $r_{200}$ (defined in the paragraph before subsection
3.1) while the ordinate is the enclosed mass ratio $M_2/M_1$ between
the hot ICM and dark matter halo. The three curves are the enclosed
mass ratio of solution $2$ with the same parameter $K_2=5.50\times
10^{20}$ in SI unit on the downstream side of a shock but with
different times as indicated along the curves.
As for different times the virial radius $r_{200}$ is a little
different, the same position in the space corresponds to different
radius in the figure in principle. However, the differences between
the virial radius at different times are so small that we can treat
one specific point of the $x$ axis as the same position in the space
at different times approximately.}
\end{figure}

\begin{figure}
\includegraphics[height=7cm]{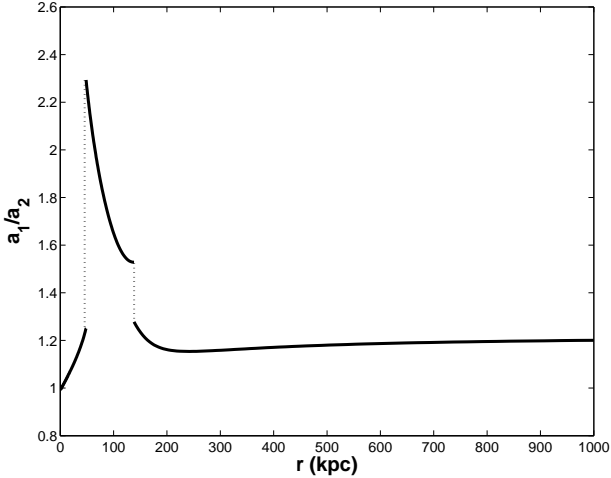}
\caption{\label{nsmall1_soundratio}
The radial profile for the polytropic sound speed ratio $a_2/a_1$ of
fluid $1$ (dark matter) to fluid $2$ (hot ICM) at time
$t=7.59\times10^{15}$ for quasi-static solution 2 shown in Figs.
$1-5$. The first jump on the left at about $r=48.38$ kpc is due to
the shock in fluid $2$ (hot ICM) and the second jump on the right at
about $r=138.23$ is due to the shock in fluid $1$ (dark matter
halo). When $x\rightarrow 0^{+}$, this ratio is about
${a_1}/{a_2}=0.99$ and when $x\rightarrow +\infty$ this ratio is
about ${a_1}/{a_2}=1.216$.}
\end{figure}

According to expression (\ref{T}), the temperature of hot ICM on the
downstream side of the shock is $\sim 0.85$keV while the temperature
on the upstream side is $\sim 0.49$ keV. The enclosed mass of hot
gas at 0.5Mpc is about $M_2=2.4\times10^{12}M_{\odot}$; the enclosed
mass ratio of dark matter halo to hot gas at 0.25Mpc is $\sim 7.58$
and at 0.5Mpc is $\sim 7.47$. By expression (\ref{vshock}), the
outgoing shock speed in the hot ICM is $u_{s,2}=nr_{s,2}/t=157.6$ km
s$^{-1}$. For a shock located at a radius of 48.38 kpc with a
reference timescale of $t=2.4\times10^{8}$ yr,
the shock position by equation (\ref{shockposition}) is
$r_{s,2}=9.54\times10^{-6}\ t^{0.8}$ where $r_{s,2}$ is in the unit
of kpc and time $t$ is in the unit of year. For example, during a
timescale of $\sim 10^{10}$ yr, this shock would travel to a radius
of $\sim 1$Mpc within a cluster of galaxies.
The enclosed mass ratio $M_2/M_1$ between the hot ICM and the dark
matter of this solution at different radii is shown in Fig.
\ref{nsmall1_massratio} (the solid line with time $t=2.4\times10^8
$ yr). If we adjust the timescale, the solution will evolve in the
self-similar manner. We choose two other timescales as examples
with other parameters of solution $2$ fixed and the result is also
shown in Fig. \ref{nsmall1_massratio}. Note that with increasing
time, the minimum of this ratio moves towards larger radii which
is the result of accretion of hot gas and outflow of dark matter
around the centre. The ratio remains nearly independent of time at
the centre and at large radii.

As an example of conceptual exercise, we calculate at the reference
time $t_1=7.59\times10^{15}$ s the energies within the radial range
$\sim 138 - 150.8$ kpc undisturbed by shocks. The gravitational
potential energy is $E_{grav,o}=-2.66\times10^{60}$ erg by
expression (\ref{genergy}), the kinetic energy of fluid $1$ is
$E_{k,1,o}=4.94\times10^{57}$ erg by expression (\ref{kenergy}), the
kinetic energy of fluid $2$ is $E_{k,2,o}=4.26\times10^{56}$ erg by
expression (\ref{kenergy}), the thermal energy of fluid $1$ is
$E_{th,1,o}=3.43\times10^{60}$ erg by expression (\ref{thenergy}),
and the thermal energy of fluid $2$ is
$E_{th,2,o}=2.72\times10^{59}$ erg by expression (\ref{thenergy}).
According to equation (\ref{energy}), the total energy is then
$E_{total,o}=1.05\times10^{60}$ erg. As $r_{s,2}=9.54\times10^{-6}\
t^{0.8}$, the two shocks have passed through the radius $150.8$ kpc
by the time $t_2=3.14\times10^{16}$ s. Then at this time $t_2$, the
energies within the same radial range are: the gravitational energy
$E_{grav,f}=-2.55\times10^{60}$ erg by expression (\ref{genergy}),
the kinetic energy of fluid $1$ $E_{k,1,f}=1.0\times10^{56}$ erg by
expression (\ref{kenergy}), the kinetic energy of fluid $2$
$E_{k,2,f}=8.32\times10^{56}$ erg by expression (\ref{kenergy}), the
thermal energy of fluid $1$ $E_{th,1,f}=3.41\times10^{60}$ erg by
expression (\ref{thenergy}), and the thermal energy of fluid $2$
$E_{th,2,f}=3.09\times10^{59}$ erg by expression (\ref{thenergy}).
According to equation (\ref{energy}), the total energy is then
$E_{total,f}=1.16\times10^{60}$ erg. Then according to equation
(\ref{power}), the mean power of shock flow in this example is
${\cal P}_{total}=4.94\times10^{42}$ erg $\hbox{s}^{-1}$.

Around a distance of $r=136$ kpc and with increasing $r$, the radial
flow velocity of hot gas changes from inflow to outflow, where we
can calculate the total mass accretion rate. We take a position
$r_a=135.98$ kpc and the mass accretion rate there is then
$\dot{M}_{a,2}=4\pi\rho_2u_2 r_a^2\approx 27.1 M_{\odot}$ yr$^{-1}$,
which is comparable to the mass accretion rates inferred for galaxy
clusters A3158(P) and A262(P). There is a summary of these
parameters for different clusters of galaxies, which are grossly
consistent with our numerical example illustrated here (e.g., Peres
et al 1998.)

At a radial distance of $\sim 3$Mpc, the outflow mass per year of
hot ICM is $\sim\dot{M}_{2,o}=4\pi\rho_2u_2r^2=211M_{\odot}$
yr$^{-1}$. Thus the approximate total outflow mass of hot gas within
a timescale of $\sim 2.4\times10^8$ yr is
$5.1\times10^{10}M_{\odot}$. Meanwhile, the inflow mass per year of
dark matter is $\sim\dot{M}_{1,o}=4\pi\rho_1u_1r^2=140M_{\odot}$
yr$^{-1}$. Then the total inflow mass of dark matter within a
timescale of $2.4\times10^8$ yr is approximately $\sim
3.36\times10^{10}M_{\odot}$. Here, we take the timescale of galaxy
cluster winds to be $\sim 6\times10^9$ yr. Then during this time, we
assume for simplicity that there is an AGN in a cluster of galaxies
every $2.4\times10^8$ yr and the hot gas mass of $\sim
5.1\times10^{10}M_{\odot}$ is lost and the dark matter mass of $\sim
3.36\times10^{10}M_{\odot}$ is deposited each time. (That means
there have been roughly 25 times of AGN activities in galaxy
clusters since initial galaxy cluster winds formed.) Furthermore,
when galaxy cluster winds began to form, we may presume that the
baryon mass fraction $f_b$ is roughly the average cosmic baryon
fraction in the expanding universe. In this example, the total mass
of hot ICM within 3 Mpc is $6.1\times10^{12}M_{\odot}$ and the total
mass of dark matter within this same radial scale is
$4.1\times10^{13}M_{\odot}$ at present. Then the value of baryon
fraction $f_b$ at present is 0.129 while the value of baryon
fraction $f_b$ when this cluster of galaxies begins to blow a galaxy
cluster wind is 0.154. Therefore the baryon fraction $f_b$ decreases
16.4\% due to galaxy cluster winds and the inflow or contraction of
dark matter halo in this example of illustration. In this scenario,
the `missing baryons' should be found in the periphery of galaxy
clusters on much larger scales. They cool down to lower temperatures
and may not be easily detected.

\subsection{Solution Examples of $n=1$}

\begin{figure}
\includegraphics[height=6cm]{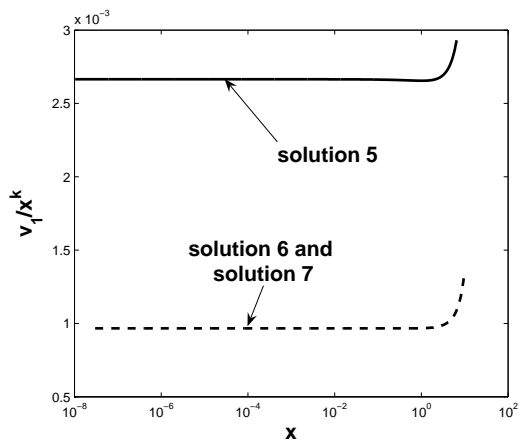}
\caption{\label{neq1_v1}Scaled radial flow velocities of fluid $1$
(dark matter) at small $x$ when $n=1$. The relevant model parameters
are summarized in Table \ref{neq1}. Note that for a certain range of
radial distance around small $x$, $v_1/x^k$ remains constant,
showing that these solutions 5, 6 and 7 for $v_1$ are indeed
quasi-static solutions as derived analytically. For all these three
solutions 5, 6 and 7, we have the same parameter set $\{n,\
\gamma_1,\ \gamma_2,\ \kappa\}=\{1,\ 1.3,\ 1.405,\ 4\}$. The
corresponding flow velocities of fluid $2$ (hot ICM) are shown in
Fig. \ref{neq1_v2}.}
\end{figure}

\begin{figure}
\includegraphics[height=6cm]{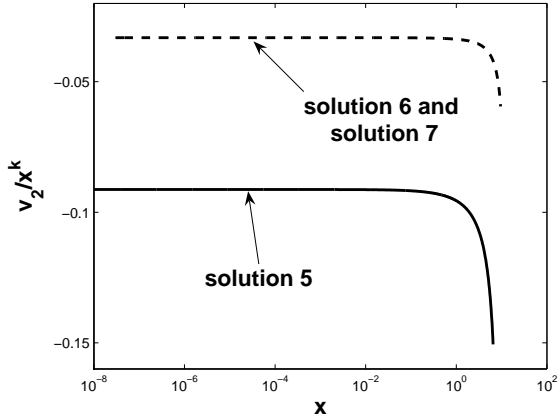}
\caption{\label{neq1_v2}Radial flow velocities of fluid $2$ (hot
gas) at small $x$ when $n=1$. The relevant model parameters are
listed in Table \ref{neq1}. Note that for a certain range of initial
distance, $v_2/x^k$ remains constant, which shows that these
solutions of $v_2$ are indeed quasi-static solutions. Together with
Fig. \ref{neq1_v1} for radial flow velocities of fluid $1$ (dark
matter halo), we have succeeded in constructing solutions with both
fluids being quasi-static at small $x$. For all these solutions, we
have the same $\{n,\ \gamma_1,\ \gamma_2,\ \kappa\}=\{1,\ 1.3,\
1.405,\ 4\}$. }
\end{figure}

\begin{figure}
\includegraphics[height=6cm]{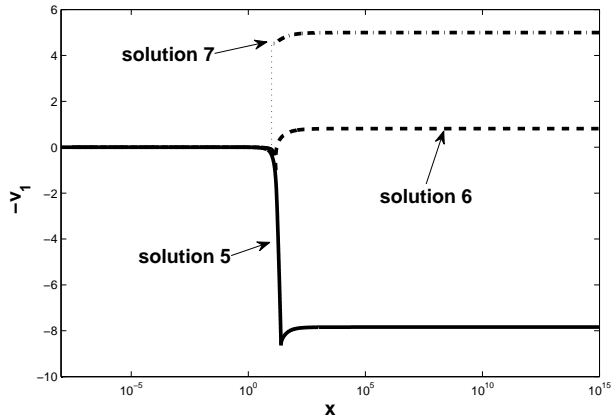}
\caption{\label{neq1_v1_2}Radial flow velocities of fluid $1$ (dark
matter halo) at large $x$ for $n=1$. The relevant model parameters
are listed in Table \ref{neq1} with $\{n,\ \gamma_1,\ \gamma_2,\
\kappa\}=\{1,\ 1.3,\ 1.405,\ 4\}$. As shown in this figure, $v_1$
remains constant for these three solutions at large $x$, indicating
that the $x_{max}$ we adopt is large enough for the asymptotic
behaviour of $v_1(x)$. The radial flow velocities of fluid $1$ at
small $x$ for these three solutions are shown in Fig. \ref{neq1_v1}.
The shock of solution 5 locates at the turning point $x=25.1$,
jumping from $v_1=8.645$ to $v_1=8.589$ and the shock of solution 6
locates at $x=15$, jumping from $v_1=0.99$ to $v_1=0.3$.}
\end{figure}

\begin{figure}
\includegraphics[height=6cm]{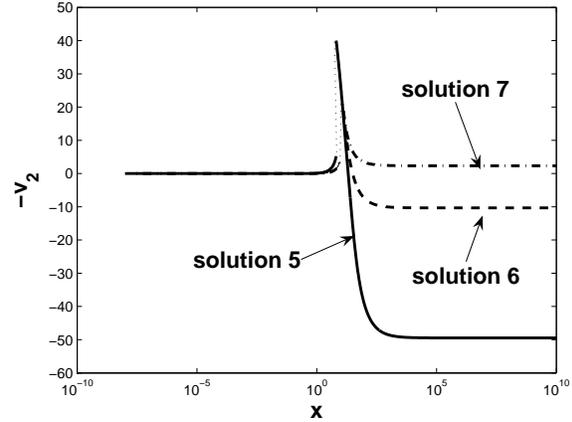}
\caption{\label{neq1_v2_2}
Radial flow velocities of fluid $2$ at large $x$
when $n=1$. The relevant model parameters are listed in Table
\ref{neq1} with $\{n,\ \gamma_1,\ \gamma_2,\ \kappa\}=\{1,\ 1.3,\
1.405,\ 4\}$. As shown in this figure, $v_2$ remains constant for
these three solutions at large $x$, indicating that the $x_{max}$ we
adopt is sufficiently large for the asymptotic behaviour of
$v_2(x)$. For solutions 6 and 7, the shock positions are the same in
gas but are different in dark matter halo (see Fig.
\ref{neq1_v1_2}). Together with Fig. \ref{neq1_v1_2}, we have
succeeded in constructing the semi-complete solutions with $n=1$.
The radial flow velocities of fluid $2$ at small $x$ for these
solutions are shown in Fig. \ref{neq1_v2}.}
\end{figure}

When $n=1$, the radial flow velocities of both fluids approach
constant values as $x\rightarrow +\infty$; for global semi-complete
solutions, these values can be negative, zero or positive at large
$x$, depending on the choice of relevant parameters and positions of
shocks. With $n=1$, the asymptotic constant speeds are $H_1$ and
$H_2$ in asymptotic velocity solutions (\ref{largev1}) and
(\ref{largev2}) respectively. Physically, these correspond to
inflows, breezes or contractions, and outflows, respectively. As
$x\rightarrow 0^{+}$, flow solutions can be matched with the
asymptotic quasi-static solution for both fluids. For $n=1$, the
mass density profile is $\rho_i\rightarrow r^{-2}$ either as
$r\rightarrow 0^{+}$ or as $r\rightarrow +\infty$. We now provide a
few examples with parameters $\{n,\ \gamma_1,\ \gamma_2,\
\kappa\}=\{1,\ 1.3,\ 1.405,\ 4\}$ and corresponding values of
$\{A_1,\ A_2,\ k\}$ being $\{267.2,\ 38.4,\ 1.89\}$. Similar to the
case of $2/3<n<1$ as described in the previous subsection, we start
the numerical integration with a small enough $x_{ini}$ and go to a
large enough $x_{max}$. Once $v_1(x)$ and $v_2(x)$ become constant
after numerical integrations in a certain small $x$ range, we would
regard the $x_{max}$ as sufficiently large. The results are shown in
Fig. \ref{neq1_v1} to Fig. \ref{neq1_v2_2} and the relevant model
parameters are summarized in Table \ref{neq1}.
Specifically for these solutions 5, 6 and 7, the nine parameters to
determine a dimensionless solution are: $\{n,\ \gamma_1,\ \gamma_2,\
\kappa,\ L_1,\ L_2,\ x_{ini},\ x_{d,1},\ x_{d,2}\}=\{1,\ 1.3,\
1.405,\ 4,$ $2.66\times10^{-3},\ -0.0913,$ $\ 1\times10^{-8},\
25.1,\ 6.5\}$ for solution $5$; $\{n,\ \gamma_1,\ \gamma_2,\
\kappa,\ L_1,\ L_2,\ x_{ini},\ x_{d,1},\ x_{d,2}\}$=$\{1,\ 1.3,\
1.405,\ 4,$ $9.67\times10^{-4},\ -0.0331,\ 3\times10^{-8},\ 15,\
9.5\}$ for solution $6$; and $\{n,\ \gamma_1,\ \gamma_2,\ \kappa,\
L_1,\ L_2,\ x_{ini},\ x_{d,1},\ x_{d,2}\}$=$\{1,\ 1.3,\ 1.405,\ 4,$
$9.67\times10^{-4},\ -0.0331,\ 3\times10^{-8},\ 12,\ 9.5\}$ for
solution $7$.
\begin{table}
\begin{center}
\caption{Model parameters for solutions in Fig. \ref{neq1_v1} to
Fig. \ref{neq1_v2_2} with $n=1$. Here, No. on the left gives the
numeric label to distinguish different solutions. Type 1 is the type
of solutions for fluid $1$ (dark matter halo) when
$x\rightarrow+\infty$ and Type 2 is the type of solutions for fluid
$2$ (hot ICM) when $x\rightarrow+\infty$. Here, $x_{d,1}$ and
$x_{d,2}$ are the independent self-similar variables on the
downstream sides of the respective shock positions for fluids $1$
and $2$.
$E_1$ is the coefficient of $\alpha_1$ and $E_2$ is the coefficient
of $\alpha_2$ at a sufficiently large $x$ (see asymptotic solution
\ref{largea1} and \ref{largea2}). $H_1$ is the velocity of fluid $1$
at a sufficiently large $x$ and $H_2$ is the velocity of fluid $2$
at a sufficiently large $x$ (see equations \ref{largev1} and
\ref{largev2}). As this table is too long horizontally, we break
this table into two parts and stack them together.} \label{neq1}
\begin{tabular}{|c|c|c|c|c|c|c|c}
 \hline
 No. & Type 1 & Type 2 & $L_1$ & $L_2$ & $x_{ini}$ \\
 \hline\hline
 5 & outflow & outflow & $2.66\times10^{-3}$ & $-0.0913$ & $1\times10^{-8}$ \\
 \hline
 6 & inflow & outflow & $9.67\times10^{-4}$ & $-0.0331$ & $3\times 10^{-8}$  \\
 \hline
 7 & inflow & inflow & $9.67\times10^{-4}$ & $-0.0331$ & $3\times 10^{-8}$ \\
 \hline\hline
  $x_{d,1}$ & $x_{d,2}$  & $E_1$ & $E_2$ & $H_1$ & $H_2$ \\
 \hline\hline
  25.1 & 6.5 & 487 & 33.3 & 7.84 & 49.5  \\
 \hline
  15 & 9.5 & 265 & 30.5 & $-0.811$ &  10.34  \\
 \hline
  12 & 9.5 & 200 &28.2 & $-5.00$ & $-2.34$  \\
\hline
 \end{tabular}
 \end{center}
\end{table}

\begin{figure}
\includegraphics[height=6cm]{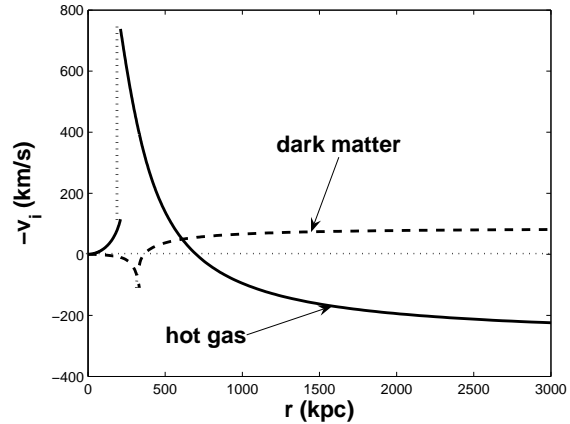}
\caption{\label{neq1_v_3}Radial flow velocities of hot ICM (solid
curve) and dark matter halo (dashed curve) of solution 6 with $n=1$,
$t=6.32\times 10^{15}$ and $K_2=7.34\times 10^8 $ SI unit in the
downstream side of a shock in hot ICM. The shock positions in hot
ICM and dark matter halo are at $\sim 210.75$ kpc and at $\sim
332.76$ kpc. The initial parameters for numerical integration are
listed in Table \ref{neq1} (solution 6). Details of the
dimensionless velocities of this solution are shown in Figs.
\ref{neq1_v1}, \ref{neq1_v2}, \ref{neq1_v1_2}, \ref{neq1_v2_2}
(solution 6).}
\end{figure}

\begin{figure}
\includegraphics[height=6cm]{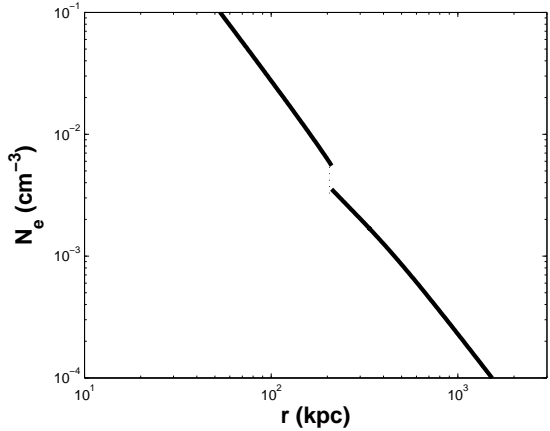}
\caption{\label{neq1_d_3}The radial profile of electron number
density with a timescale of $t=6.32\times 10^{15}$ s and
$K_2=7.34\times 10^8$ SI unit on the downstream side of a shock in
the hot ICM for solution 6 of $n=1$. Here, we also take the typical
mean molecular weight in clusters of galaxies to be $\sim 0.59$
g/mol. The radial flow velocities of this solution 6 are shown in
Fig. \ref{neq1_v_3}. }
\end{figure}

\begin{figure}
\includegraphics[height=6cm]{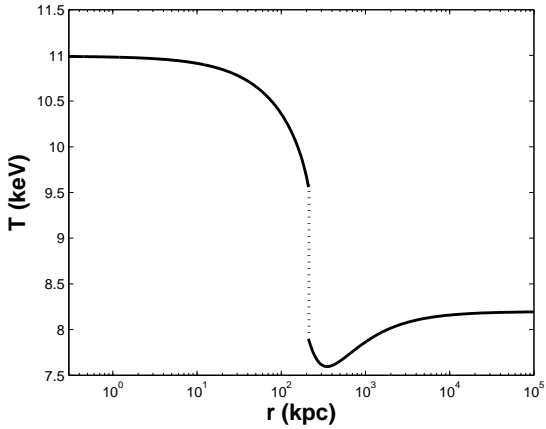}
\caption{\label{neq1_T_3}The radial temperature profile of hot ICM
with a timescale of $t=6.32\times 10^{15}$ s and $K_2=7.34\times
10^8 $ SI unit
on the downstream side of a shock in ICM for solution 6 of $n=1$.
During a range of short radial distance from the centre, the
temperature is a constant of $\sim 11$ keV and at large $x$ the
temperature is $\sim 8.2$ keV. The electron number density of this
solution 6 is shown in Fig. \ref{neq1_d_3} and the corresponding
radial flow velocities are shown in Fig. \ref{neq1_v_3}.}
\end{figure}

\begin{figure}
\includegraphics[height=7cm]{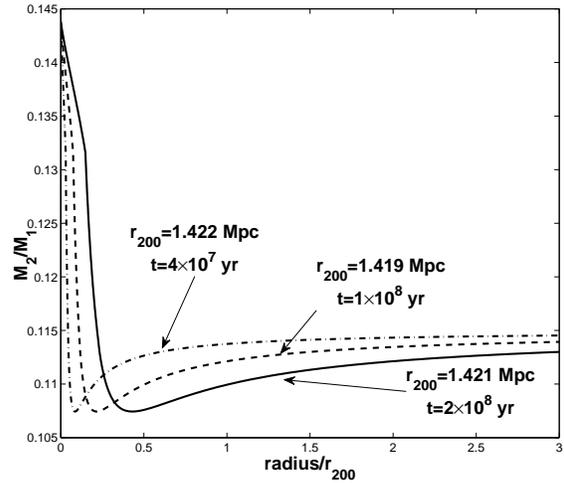}
\caption{\label{neq1_massratio}
The radial profile of enclosed mass ratio $M_2/M_1$ at different
times of solution 6 shown in Figs. $7-13$. The abscissa is the
radius in unit of virial radius $r_{200}$ and the ordinate is the
enclosed mass ratio $M_2/M_1$ between the hot ICM and dark matter.
The three curves are enclosed mass ratio of solution 6 with the same
parameter $K_2=7.34\times10^{8}$ SI unit on the downstream side of
an ICM shock and three different times $t$ marked along the curves.
As for different times the virial radius $r_{200}$ is a somewhat
different, the same position in space corresponds to different radii
in this figure. However, these differences among the virial radii at
different times are so small that we may treat one specific point of
the abscissa as the same position in space at different times
approximately.}
\end{figure}

\begin{figure}
\includegraphics[height=7cm]{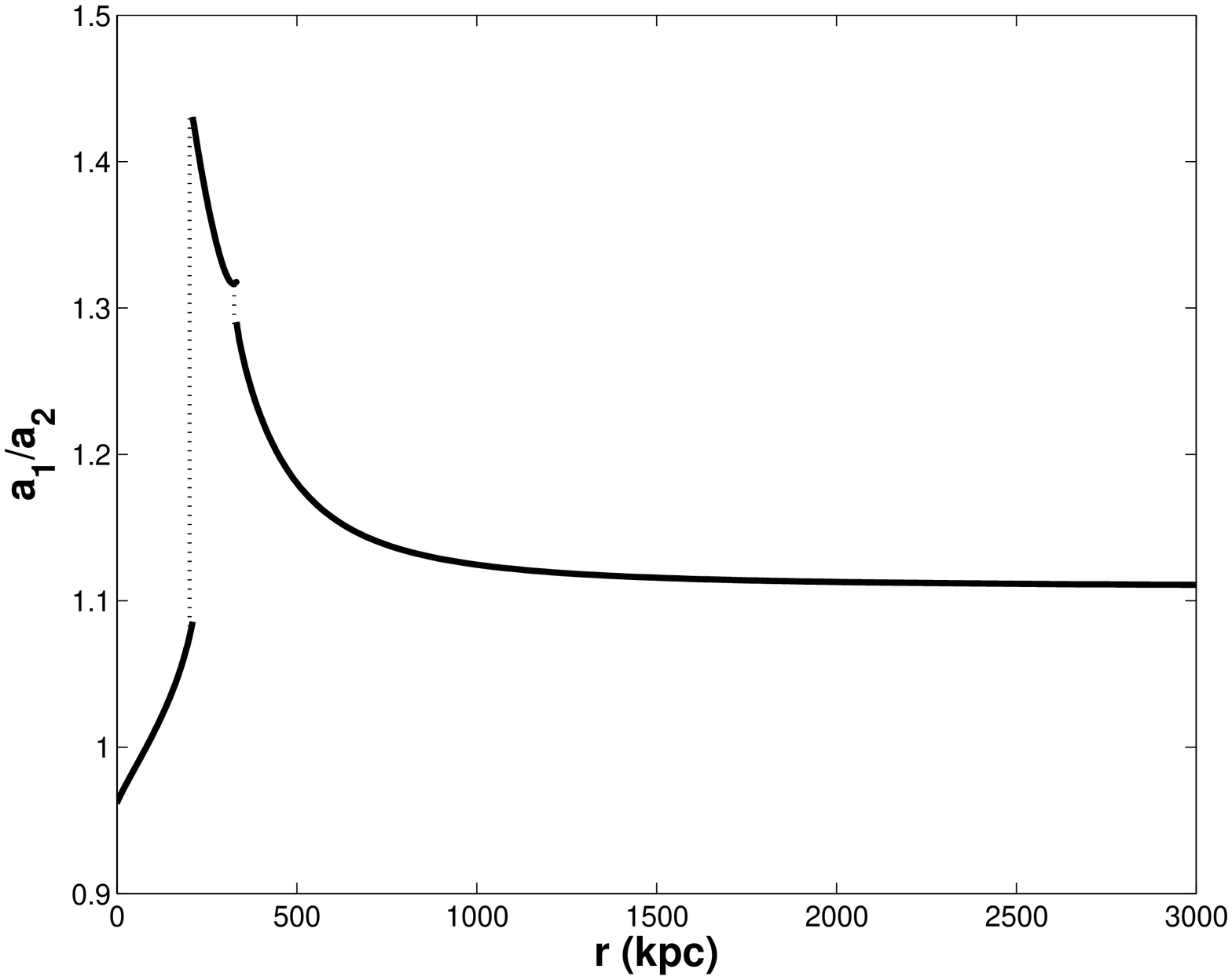}
\caption{\label{neq1_soundratio}
The radial profile for the ratio $a_1/a_2$ of polytropic sound
speeds in fluid $1$ (dark matter) to fluid $2$ (hot ICM) at time
$t=6.32\times 10^{15}$ for quasi-static solution 6 shown in Figs.
$7-14$. The first jump on the left at about $r=210.75$ kpc is due to
the shock in fluid $2$ and the second jump on the right at about
$332.76$ kpc is due to the shock in fluid $1$. When $x\rightarrow
0^{+}$, this sound speed ratio is about ${a_1}/{a_2}=0.962$ and when
$x\rightarrow +\infty$, this ratio is about ${a_1}/{a_2}=1.11$.}
\end{figure}

Again, we use solution 6 to further illustrate the process of a
galaxy cluster wind. Here, fluid $1$ represents the dark matter halo
and fluid $2$ represents the hot ICM. We choose the parameters $t$
and $K_2$ (the value on the downstream side of an ICM shock) as
$t=6.32\times 10^{15}$ s and $K_2=7.34\times 10^8$ SI unit. In this
example, the shock positions in the hot ICM and in dark matter halo
are $r_{s,2}=210.75$ kpc and $332.76$ kpc, respectively. In galaxy
cluster Hydra A, the observed shock is located at $\sim 211$ kpc
(e.g., Nulsen et al. 2005b) and the shock in galaxy cluster
Ms0735.6+7421 is at $\sim 240$ kpc (e.g., McNamara et al. 2005). For
this example, we apply our model to the radial domain of $\sim 1$
Mpc (some rich clusters of galaxies may have radial size scales of
$\sim 1-2$ Mpc; e.g., Bahcall 1996). The radial flow velocity
profiles are shown in Fig. \ref{neq1_v_3} while the profiles of
electron number density and ICM temperature are displayed in Figs.
\ref{neq1_d_3} and \ref{neq1_T_3}, respectively. Actually the
temperature profile similar to Fig \ref{neq1_T_3} in the entire
radial range has been observed in galaxy cluster NGC1275 (e.g.,
Churazov et al. 2003).
The ratio of polytropic sound speeds (see Appendix G) in two fluids
in this example is shown in Fig. \ref{neq1_soundratio}. In this
example, the radial outflow velocity of hot ICM at 1Mpc is $\sim
97.4$ km s$^{-1}$ and the radial inflow velocity of dark matter halo
at 3 Mpc is $\sim 66.76$ km s$^{-1}$. The temperature of hot ICM
downstream of the shock is $\sim 9.56$ keV and that of the upstream
of the shock is $\sim 7.9$ keV. The enclosed mass of hot ICM at
0.5Mpc is about $M_2=4\times10^{13}M_{\odot}$; and the enclosed mass
ratio of dark matter halo to hot ICM at 0.25Mpc is $\sim 8.1$ and at
0.5Mpc is $\sim 9.3$. The travel speed of the outgoing shock in the
hot ICM is $u_{s,2}=n r_{s,2}/t=1.03\times 10^3$ km s$^{-1}$. For a
shock located at 210.75 kpc with a timescale of $t=2\times 10^{8}$
yr, the shock position (see expression (\ref{shockposition})) can be
calculated by $r_{s,2}=1.05\times10^{-6}t$ where $r_{s,2}$ is in the
unit of kpc and time $t$ is in the unit of year. For $n=1$, the
shock speed remains constant.
The enclosed mass ratio $M_2/M_1$ between the hot ICM and the dark
matter of this solution 6 shown Figs. $7-13$ at different radii is
displayed in Fig. \ref{neq1_massratio} (the solid line with time
$t=2\times10^8 $ yr). If we adjust the timescale $t$, solution 6
will evolve in a self-similar manner. We choose two other timescales
$t=10^8$ yr (dashed curve) and $4\times 10^7$ yr (dash dotted curve)
as two additional examples with other parameters of solution $6$
unchanged and the results are shown in Fig. \ref{neq1_massratio}.
With increasing time $t$, the minimum of this enclosed mass ratio
moves towards larger radii which is the result of accretion of hot
ICM and outflow of dark matter around the centre. This enclosed mass
ratio is nearly independent of time $t$ around the centre and at
large radii.

As another example of conceptual exercise, we calculate at a
reference time $t=6.32\times 10^{15}$ s the energies within radial
range $\sim 332.7 - 502$ kpc undisturbed by shocks below. The
gravitational energy is $E_{grav,o}=-9.74\times10^{63}$ erg (see
equation \ref{genergy}), the kinetic energy of fluid $1$ is
$E_{k,1,o}=7.39\times10^{59}$ erg (see equation \ref{kenergy}), the
kinetic energy of fluid $2$ is $E_{k,2,o}=8.42\times10^{60}$ erg
(see equation \ref{kenergy}), the thermal energy of fluid $1$ is
$E_{th,1,o}=1.97\times10^{64}$ erg (see equation \ref{thenergy}),
the thermal energy of fluid $2$ is $E_{th,2,o}=1.05\times10^{63}$
erg (see equation \ref{thenergy}). According to equation
(\ref{energy}), the total energy is then given by
$E_{total,o}=1.11\times10^{64}$ erg. As
$r_{s,2}=1.05\times10^{-6}t$, the two shocks have passed through the
radius $502$ kpc by time $t_2=1.50\times10^{16}$ s. Then at this
time the energies within the same radial range are: the
gravitational energy $E_{grav,f}=-1.01\times10^{64}$ erg (see
equation \ref{genergy}), the kinetic energy of fluid $1$
$E_{k,1,f}=5.78\times10^{58}$ erg (see equation \ref{kenergy}), the
kinetic energy of fluid $2$ $E_{k,2,f}=7.76\times10^{59}$ erg (see
equation \ref{kenergy}), the thermal energy of fluid $1$
$E_{th,1,f}=1.97\times10^{64}$ erg (see equation \ref{thenergy}),
the thermal energy of fluid $2$ $E_{th,2,f}=1.73\times10^{63}$ erg
(equation \ref{thenergy}). According to equation (\ref{energy}), the
total energy is then $E_{total,f}=1.13\times10^{64}$ erg. Then
according to equation (\ref{power}), the mean power of shock flow in
this illustration example is ${\cal P}_{total}=2.94\times10^{46}$
erg $\hbox{s}^{-1}$.

Around a radial distance of $695.5$ kpc and with increasing $r$,
the flow velocity of hot gas changes from inflow to outflow, where
we can calculate the total mass accretion rate. We take a radius
of $r_a=695$ kpc and the mass accretion rate is then
$\dot{M}_{a,2}=4\pi\rho_2u_2r_a^2\approx 193M_{\odot}$ yr$^{-1}$,
comparable to the mass accretion rates in galaxy clusters A85(P)
and A644(P) (e.g., Peres et al. 1998).

At a radius of $\sim 1$Mpc, the outflow mass per year of hot ICM is
$\sim\dot{M}_{2,o}=4\pi\rho_2u_2r^2=8.11\times10^{3}M_{\odot}$
yr$^{-1}$, which is comparable to the steady wind result of Yahil \&
Ostriker (1973). Thus the approximate total outflow mass of hot gas
within a timescale of $\sim 2\times10^8$ year is $\sim
1.6\times10^{12}M_{\odot}$. Meanwhile, the inflow mass per year of
dark matter is
$\sim\dot{M}_{1,o}=4\pi\rho_1u_1r^2=4.94\times10^{4}M_{\odot}$
yr$^{-1}$. Then the total inflow mass of dark matter within a
timescale of $2\times10^8$ yr is approximately $10^{13}M_{\odot}$.
Here, we take the timescale of galaxy cluster winds to be
$3\times10^9$ yr and the AGN activity occurs every $2\times10^8$ yr.
(That means there have been roughly 15 times of AGN activities in
galaxy clusters since the initial wind formed.) In this example, the
total mass of hot ICM within 1 Mpc is $8.1\times10^{13}M_{\odot}$
and the total mass of dark matter within this same radial scale is
$\sim 7.4\times10^{14}M_{\odot}$ at present. Now consider the mass
loss of hot ICM due to galaxy cluster winds and inflow of dark
matter. The total mass of hot ICM within this radial domain when the
galaxy cluster winds began to form was $1.05\times10^{14}M_{\odot}$
and the total mass of dark matter within this same radial domain
when the galaxy cluster winds began to form was
$5.96\times10^{14}M_{\odot}$. Therefore the baryon fraction $f_b$ at
present is 0.098 and its value when the cluster began to form winds
is 0.15. The value of $f_b$ decreases 34.7\% due to galaxy cluster
winds and inflow of dark matter halo in this example of $n=1$.

\subsection{Solution Examples of $n>1$}

With $n>1$ in our model, radial flow velocities of hot gas and dark
matter will tend to diverge for $H_1\neq 0$ and $H_2\neq 0$ as we
integrate towards $x\rightarrow +\infty$ (see asymptotic solutions
(\ref{largev1}) and (\ref{largev2}) towards large $x$) with
quasi-static asymptotic solutions around the cluster centre. This
case of $n>1$ can be relevant as the density profile scales as
$\rho\propto r^{-2/n}$ while in many clusters of galaxies the
electron number densities observed scale as a power law with power
indices between $-1$ and $-2$ (e.g., clusters of galaxies A2204,
A2052 and Ms0735.6+7421). So for a physical system, we only apply
our model to a finite size of the order of $\sim 1$ Mpc, which is
the size of a typical cluster of galaxies. Therefore, we have
considerable interest in those solutions with two coefficients $H_1$
and $H_2$ of the diverging terms being zero. We are now in a
position to show such an example of solutions with $n>1$.

\begin{figure}
  \includegraphics[height=6cm]{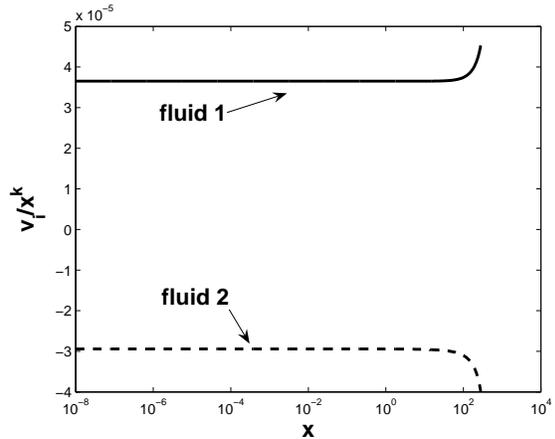}\\
  \caption{The scaled radial flow velocities as $x\rightarrow 0^+$ for convergent
solutions at large $x$ when $n>1$ and with $H_1=H_2=0$. The model
parameters at small $x$ are $\{n,\ \gamma_1,\ \gamma_2,\ \kappa,\
x_{ini},\ L_1,\ L_2\}=\{1.07,\ 1.315,\ 1.42,\ 0.1,\ 1\times10^{-8},\
3.65\times10^{-5},\ -2.94\times10^{-5}\}$. This figure shows that
the solution approaches quasi-static configuration for both fluids
with this set of parameters at small $x$. These parameters are kept
the same as we try to find the convergent solution. So the solution
we find here is still quasi-static at small $x$.}
  \label{match_v_small}
\end{figure}

\begin{figure}
  \includegraphics[height=6cm]{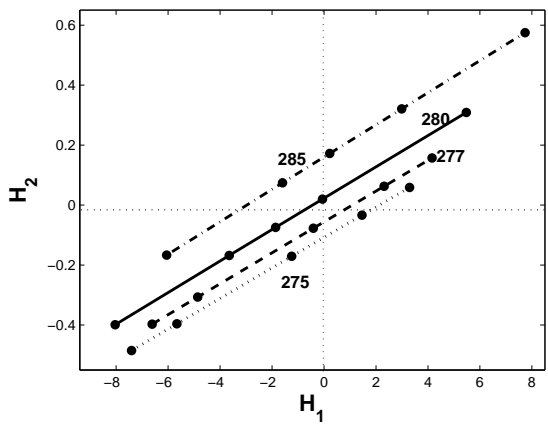}\\
  \caption{The numerical search of a global semi-complete solution
of $n>1$ with $H_1=0$ and $H_2=0$. As an example, the dash-dotted
line marked by 285 is the line with $x_{d,2}=285$. As $x_{d,1}$
increases, the point goes from the bottom left to upper right
(values of both $H_1$ and $H_2$ increase) along a straight line.
Other straight lines are plotted in the same manner with different
values of $x_{d,2}$ as explicitly marked. The solid, heavy dashed,
and dotted lines are with $x_{d,2}=280,\ 277,\ 275$, respectively.
The trend of variation is clear by this numerical exploration. The
parameters of these four straight lines are summarized in Table
\ref{table_a2204}. Other parameters at small $x$ are $\{n,\
\gamma_1,\ \gamma_2,\ \kappa,\ x_{ini},\ L_1,\ L_2\}=\{1.07,\
1.315,\ 1.42,\ 0.1,\ 1\times10^{-8},\ 3.65\times10^{-5},\
-2.94\times10^{-5}\}$.} \label{match_a2204}
\end{figure}

In our model, we choose relevant parameters to be $\{n,\ \gamma_1,\
\gamma_2,\ \kappa\}=\{1.07,\ 1.315,\ 1.42,\ 0.1\}$ with the
corresponding values of $\{A_1,\ A_2,\ k\}=\{1.53\times10^5,\
2.27\times10^4,\ 1.81\}$. The parameters at small $x$ are
$\{x_{ini},\ L_1,\ L_2\}=\{1\times10^{-8},\ 3.65\times10^{-5},\
-2.94\times10^{-5}\}$. The radial flow velocities at small $x$ are
shown in Fig. \ref{match_v_small}, confirming that the solution we
find is indeed quasi-static as $x\rightarrow 0^{+}$. By numerical
exploration, we can choose proper values of independent self-similar
variables $x_{d,1}$ and $x_{d,2}$ on the downstream sides of
respective shock positions and construct a solution with $H_1=0$ and
$H_2=0$ (i.e., finite radial flow velocities at large $x$). With the
relevant parameters chosen for small $x$, we vary the shock
positions in fluid $1$ and fluid $2$, respectively, and compute the
coefficients $H_1$ and $H_2$. When the product $v_i(x)\ x^{-1+1/n}$
approaches constant values for either values of $i$ after a certain
$x_{max}$, we would regard $x$ as being large enough for asymptotic
solutions and then evaluate the values of $H_1$ and $H_2$.
Empirically, we find that for a fixed $x_{d,2}$ value, if we change
the value of $x_{d,1}$ continuously, $H_1$ and $H_2$ lie on a
perfect straight line. When we vary $x_{d,2}$, this $H_1$ versus
$H_2$ line will move. For a certain $x_{d,2}$, this line can move
across the zero point in the functional relation of $H_1$ versus
$H_2$. When this happens, there exists a pair of $\{x_{d,1},\
x_{d,2}\}$ such that $H_1=H_2=0$. This corresponds to convergent
radial flow velocities for both fluids at large $x$. The result is
then shown in Fig. \ref{match_a2204} and the relevant parameters are
summarized in Table \ref{table_a2204}.

\begin{table}
\begin{center}
\caption{Model parameters for constructing a solution of $n>1$
with $H_1=0$ and $H_2=0$. Here, $x_{d,1}$ and $x_{d,2}$ are the
independent self-similar variables on the downstream sides of the
shock positions for fluid $1$ (dark matter) and fluid $2$ (hot ICM),
respectively. $H_1$ is the coefficient of $v_1(x)$ and $H_2$ is the
coefficient of $v_2(x)$ when $x$ becomes sufficiently large. Other
model parameters at small $x$ are $\{n,\ \gamma_1,\ \gamma_2,\
\kappa,\ x_{ini},\ L_1,\ L_2\}=\{1.07,\ 1.315,\ 1.42,\ 0.1,\
1\times10^{-8},\ 3.65\times10^{-5},\ -2.94\times10^{-5}\}$.}
\label{table_a2204}
\begin{tabular}{|c|c|c|c|}
\hline
 $x_{d,2}$& $x_{d,1}$ & $H_1$ & $H_2$ \\
 \hline\hline
 285 & 550 & $-6.056$ & $-0.1668$\\
 285 & 555 & $-1.595$ & $0.07426$\\
 285 & 557 & 0.2219 & 0.1721 \\
 285 & 560 & 2.989  & 0.3208 \\
 285 & 565 & 7.736  & 0.5748 \\
 \hline
 280 & 550 & $-8.039$ & $-0.3993$\\
 280 & 555 & $-3.647$ & $-0.1681$\\
 280 & 557 & $-1.862$ & $-0.0744$\\
 280 & 559 & $-0.0598$& 0.0199 \\
 280 & 565 & 5.4775   & 0.309  \\
 \hline
 277 & 553 & $-6.613$ & $-0.3972$ \\
 277 & 555 & $-4.860$ & $-0.3067$ \\
 277 & 560 & $-0.411$ & $-0.0774$ \\
 277 & 563 & 2.314  & 0.0625 \\
 277 & 565 & 4.1578 & 0.1569 \\
 \hline
 275 & 553 & $-7.406$ & $-0.4851$ \\
 275 & 555 & $-5.663$ & $-0.3961$ \\
 275 & 560 & $-1.242$ & $-0.171$ \\
 275 & 563 & 1.462 & $-0.0338$ \\
 275 & 565 & 3.290 & 0.0587 \\
 \hline
  \end{tabular}
  \end{center}
\end{table}

\begin{figure}
\includegraphics[height=6cm]{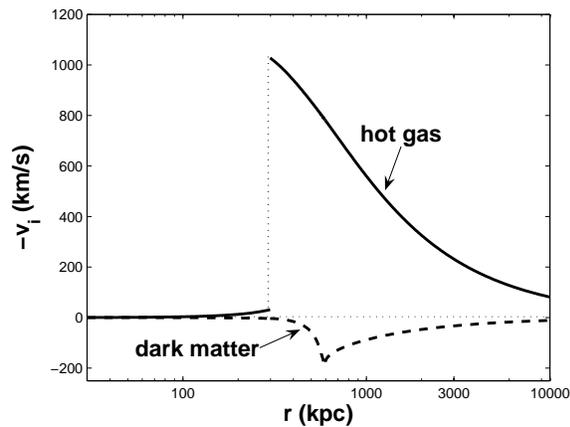}
\caption{\label{match_v}The radial flow velocities of hot ICM and
dark matter halo of convergent solution with $n>1$, $t=1.1\times
10^{16}$ s, and a downstream $K_2=5.03\times 10^6$ SI unit in the
hot ICM. The shock position in hot ICM is at $\sim 298.6$ kpc and
the shock position in the dark matter halo is at $\sim 596.2$ kpc.
The relevant parameters for the numerical integration are $\{n,\
\gamma_1,\ \gamma_2,\ \kappa,\ x_{ini},\ L_1,\ L_2\}=\{1.07,\
1.315,\ 1.42,\ 0.1,\ 1\times10^{-8},\ 3.65\times10^{-5},\
-2.94\times10^{-5}\}$. }
\end{figure}

\begin{figure}
\includegraphics[height=6cm]{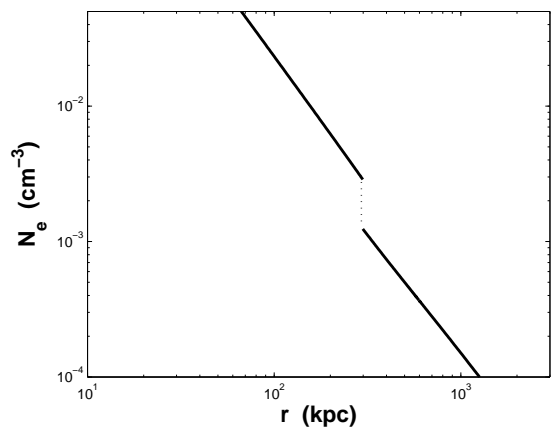}
\caption{\label{match_d}The electron number density profile of a
convergent solution of $n>1$ with $t=1.1\times10^{16}$ s and
$K_2=5.03\times 10^6$ in SI unit in the downstream side of the shock
in the hot ICM. Here, we also take the typical mean molecular weight
in galaxy clusters to be $0.59$ g/mol (e.g., Cavaliere \&
Fusco-Femiano 1978). The other parameters are $\{n,\ \gamma_1,\
\gamma_2,\ \kappa,\ x_{ini},\ L_1,\ L_2\}=\{1.07,\ 1.315,\ 1.42,\
0.1,\ 1\times10^{-8},\ 3.65\times10^{-5},\ -2.94\times10^{-5}\}$. }
\end{figure}

\begin{figure}
\includegraphics[height=6cm]{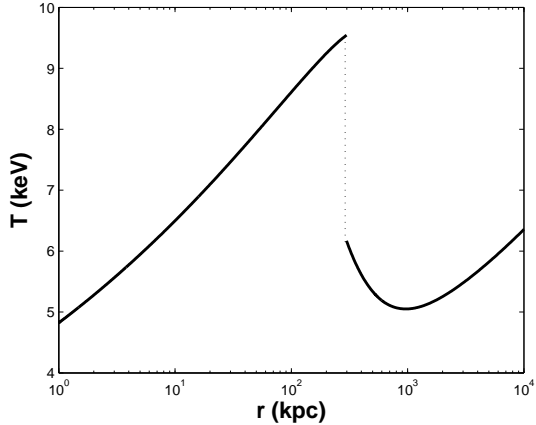}
\caption{\label{match_T}The temperature profile of hot ICM for a
convergent solution at large $x$ of $n>1$ with $t=1.1\times10^{16}$
s and $K_2=5.03\times 10^6$ in SI unit in the downstream side of the
shock in the hot ICM. The temperature increases with increasing
radius at both small and large $x$. }
\end{figure}
From Fig. \ref{match_a2204}, we know the existence of a solution
with convergent flow velocities at large $x$ with $x_{d,2}=280$ and
$x_{d,1}=559$. We shall describe the physical process that this
solution represents presently. The time $t$ is chosen to be
$t=1.1\times10^{16}$ s and the downstream $K_2$ of hot ICM is chosen
to be $K_2=5.03\times10^6$ SI unit. Then the shock position in hot
ICM is at $\sim 298.6$ kpc and the shock in the dark matter halo is
at $\sim 596.2$ kpc.

We apply this numerical example to within a radial scale of $\sim 3$
Mpc. The radial flow velocity profiles are shown in Fig.
\ref{match_v}. If the two coefficients $H_1$ and $H_2$ are exactly
zero, the hot gas velocity is inward and vanishes, while the flow
velocity of dark matter is outward and vanishes as $r\rightarrow
+\infty$. The profile of electron number density is shown in Fig.
\ref{match_d} and the temperature profile is shown in Fig.
\ref{match_T}. In this example, the inflow velocity of hot gas at
3Mpc is about $231.4$ km s$^{-1}$ while the outflow velocity of dark
matter at 3Mpc is $\sim 32.8$ km s$^{-1}$. Across the shock, the
temperature of hot ICM on the downstream side is $\sim 9.5$ keV and
that on the upstream side is $\sim 6.17$ keV. The enclosed mass of
hot ICM at 0.5Mpc is about $M_2=3.33\times10^{13}M_{\odot}$; and the
enclosed mass ratio of dark matter to the hot ICM at 0.25Mpc is
$\sim 7$ and at 0.5Mpc is $\sim 9.5$. The outgoing shock speed in
the hot ICM is $u_{s,2}=nr_{s,2}/t=892$ km s$^{-1}$. For a shock
located at $r=298.6$ kpc with a timescale of $t=3.8\times10^{8}$ yr,
the shock position given by expression (\ref{shockposition}) can be
followed by $r_{s,2}=2.2\times10^{-7}\ t^{1.07}$ where $r_{s,2}$ is
in the unit of kpc and time $t$ is in the unit of year.

\section{Summary}

In this paper, we formulate a two-fluid model framework of spherical
symmetry to explore dynamic behaviours of hot ICM and dark matter
during the evolution of galaxy clusters. In this scenario, the hot
ICM and dark matter halo are approximated as two polytropic `fluids'
and are coupled by gravity. For both `fluids' in general situations,
specific entropies are conserved along streamlines separately and
are related to the enclosed masses. Quasi-static solutions for both
`polytropic fluids' can be obtained and are adopted for sufficiently
small radii around the central core region at a given time. In order
to construct dimensionless quasi-static solutions in the regime of
small $x$, we need to specify nine dimensionless parameters $\{n,\
\gamma_1,\ \gamma_2,\ \kappa,\ L_1,\ L_2,\ x_{ini},\ x_{d,1},\
x_{d,2}\}$ where $x_{ini}$ needs to be properly chosen. Two
additional dimensional parameters $K_2$ and $t$ need to be specified
for dimensional solutions for a physical description. The ICM
temperature is taken to be on the order of $\sim 10^7-10^8$ K in
typical clusters of galaxies and the electron number density is
taken to be $\sim 10^{-2}-10^{-4}$ cm$^{-3}$ in the typical radial
range of kpcs to Mpcs. It is possible to construct different types
of flow solutions with shocks in both fluids and with various
asymptotic scaling features in flow speeds, mass densities, enclosed
mass and temperatures. In particular, we can construct dynamic
solutions for galaxy cluster winds and discuss the important problem
of `missing baryons' during the evolution of galaxy clusters. There
are several physical hypotheses in our two-fluid model to simplify
the mathematical analysis. First, we take the dark matter halo as a
kind of `fluid' for simplicity and explore this alternative
theoretical possibility. Secondly, we assume the two-fluid system of
galaxy clusters to be grossly spherically symmetric with a common
centre and hope to catch major dynamic flow features on large
scales. Thirdly, both `fluids' are assumed to be `polytropic' in the
most general sense of entropy conservation along streamlines. In
other words, the specific entropy distribution is not necessarily
constant in space and time but are allowed to vary in $r$ and $t$ in
general. Finally, dark matter interacts with the hot ICM only
through gravity.

Global semi-complete solutions can be constructed to pass through
the two singular surfaces via shocks in both fluids. As large-scale
ICM shocks have been identified observationally, there must be
large-scale flows of hot ICM. As the dark matter halo and the hot
ICM are coupled by gravity, there may be dark matter flows and the
possibility of shocks in dark matter halo. Such dark matter shocks
are characterized by drastic density jumps and sharp rises of
velocity dispersions and may be detected by utilizing gravitational
lensing effects. In our model framework, outflows of hot ICM in
galaxy clusters actually form galaxy cluster winds, which is a
systematic mechanism of reducing the baryon fraction $f_b$. In our
model, the self-similar shocks travel outwards in both hot ICM and
dark matter halo, respectively. The travel speeds of these shocks
actually change with time for $n\neq 1$ and their radial positions
at present can vary from tens of kpcs to a few Mpc. Due to galaxy
cluster winds, the baryon fraction $f_b$ in galaxy clusters can be
$\sim 15\%-40\%$ lower than the average cosmic baryon fraction in
the Universe, which may account for the problem of `missing baryons'
in clusters of galaxies. Physically, these `missing baryons' should
reside in the periphery of galaxy clusters in the form of warm gas
as results of unavoidable radiative cooling. Since the lower baryon
fraction $f_b$ is a generic phenomenon in clusters of galaxies, we
therefore suggest that galaxy cluster winds would be common and
frequent during the evolution of galaxy clusters.

The main features of our self-similar polytropic solution for two
fluids coupled by gravity are summarized below. The radial profile
of mass density at a given time is $\rho_i\propto r^{-2/n}$ in both
fluids for either $r\rightarrow 0^{+}$ or $r\rightarrow+\infty$ with
$2/3<n<2$ in general. That shows that the asymptotic mass density
profiles of hot ICM and dark matter take the same form of power laws
with the same index $-2/n$. Meanwhile, the radial flow velocities of
both fluids approach zero as $x\rightarrow 0^{+}$ in the form of
quasi-static asymptotic solution (\ref{quasiv}) and (\ref{quasia}).
At large $x$, the asymptotic flow solution is (\ref{largea1}),
(\ref{largea2}), (\ref{largev1}), (\ref{largev2}) for $2/3<n<2$.
When $2/3<n<1$ and $x\rightarrow+\infty$, the asymptotic radial flow
velocities of both fluids approach zero. When $n=1$ and
$x\rightarrow+\infty$, the radial flow velocities of both fluids
approach constant values, which may be positive, zero or negative
for various combinations of two fluids. When $n>1$ and
$x\rightarrow+\infty$, the radial flow velocities of both fluids
become divergent for $H_1\neq 0$ and $H_2\neq 0$ in asymptotic
solution (\ref{largev1}) and (\ref{largev2}). For $H_1=H_2=0$, the
radial flow velocities of both fluids remain finite at large $x$ for
$n>1$. The radial profile of temperature in the hot ICM is
$T_2\propto r^{2-2/n}$ for either $r\rightarrow 0^{+}$ or
$r\rightarrow+\infty$. By mass conservation, the enclosed mass is
continuous across shocks.

\section{Discussion}

There are several relevant aspects that may be further taken into
account in our theoretical model development. First, we have ignored
the magnetic field permeated in the hot ICM which has been
observationally inferred to range from a few $\mu$G to several tens
of $\mu$G in the central region (e.g., Fabian 1994; Carilli \&
Taylor 2002; Hu \& Lou 2004). Cluster magnetic field can affect the
gas dynamics in a nontrivial manner and can provide valuable
diagnostic information such as synchrotron emission and modified
Sunyaev-Zel'dovich effect (e.g., Hu \& Lou 2004; Bagchi et al.
2006). To better understand behaviours of a magnetized hot gas and
dark matter coupled by gravity, a magnetohydrodynamic (MHD) approach
may be adopted (e.g., Lou \& Wang 2007; Wang \& Lou 2007). Secondly,
we have assumed the flow system to be spherically symmetric, which
is a highly idealized situation. Various instabilities may arise in
dynamic flows to destroy spherical symmetry (Lou \& Bai, 2007 in
preparation). For systems of merging galaxies or galaxy clusters,
spherical symmetry is only a gross approximation. Thirdly, our
current model requires that the mass density profile has the same
form of power law scalings either at small or large radii. If the
scaling parameter $n$ can be adjusted in some proper way across a
shock, this model may be adapted to simulate more diverse kinds of
galaxy clusters. Finally, it is possible to take a distribution
function approach for dark matter halo and fluid description for the
hot ICM to model a galaxy cluster as a further development.
Admittedly, this approach could be mathematically challenging.

Compared with previous results, we note a few points regarding our
model. First, our dynamic model of galaxy cluster wind differs from
that of Yahil \& Ostriker (1973). They considered a steady-state
wind while the galaxy cluster wind of our model is dynamic and
self-similar. Their gravitational potential is static and they did
not model shock features occurred in galaxy clusters. Secondly, the
case of $n=1$ for a single fluid sphere was discussed in Suto \&
Silk (1988). In addition to the major difference of two fluids
coupled by gravity, our model with $n=1$ differs from that of Suto
\& Silk (1988) in that we have specific entropy conservation along
streamlines. In Lou (2005), the model of two isothermal fluids
coupled by gravity is discussed. Here, we treat two polytropic
fluids coupled by self-gravity in more general situations. In
particular, the quasi-static solution does not exist in an
isothermal fluid (Lou \& Wang 2006). By choosing different values of
parameters $n$, $\gamma_i$, and $\kappa$ and shock positions, our
polytropic model can be adapted to various astrophysical systems,
including clusters of galaxies and globular clusters on much smaller
scales.

Due to possible galaxy cluster winds, baryons in galaxy clusters can
flow out of galaxy clusters and stay in their periphery. Meanwhile,
dark matter halo contracts under gravity. As time goes on, such
baryon gas cools down and may not be detectable in X-ray bands. This
may naturally explain the problem of `missing baryons' for clusters
of galaxies. With this scenario in mind, we should develop
observational diagnostics to look for signatures of these missing
baryons around clusters of galaxies on much larger spatial scales
($\gsim$ Mpcs). In our perspective, the giant radio arcs recently
discovered at 1.4GHz in galaxy cluster Abell 3376 (Bagchi et al.
2006) and the strikingly similar radio arcs in galaxy cluster 3667
(e.g., Rottgering et al. 1997; Roettiger et al. 1999) are most
likely large-scale magnetohydrodynamic (MHD) shocks and imply
large-scale flows around clusters of galaxies (Lou \& Jiang 2008 in
preparation).

As mentioned earlier, AGN activities are directly observed through
intense electromagnetic radiation from normal matter. For AGN
activities in the central region of a galaxy cluster, accretion of
dark matter onto a SMBH may be involved (e.g., Hu et al. 2006)
although dark matter activities cannot be directly detected. In this
scenario, both ICM and dark matter coupled by gravity are active
components of AGN activities and violent relaxation occurs in highly
disturbed dark matter halo (Lynden-Bell 1967). In this sense, a dark
matter halo gains gravitational energy and `thermal' energy with
higher velocity dispersions. Furthermore, shocks can also form in
association with cluster merging processes also observed to occur in
clusters of galaxies (e.g., the galaxy cluster 1E 0657-56; Clowe et
al. 2006). In such merging processes, dark matter is dragged along
and the energy may be transferred to the dark matter. This mechanism
has already been discussed by some authors (e.g., Shchekinov \&
Vasiliev 2006; Knebe et al. 2002).

Very recently, caustics in dark matter halo have been discussed by
many authors (e.g., Natarajan \& Sikivie 2007; Onemli \& Sikivie
2007; Mohayaee et al. 2007). Caustics are consequence of
collisionless cold dark matter in galactic halos or clusters of
galaxies. Caustics are regions of infinite density in the limit that
the DM particles have zero velocity dispersion (e.g., Natarajan \&
Sikivie 2007). Gravitational lensing effects have been proposed to
detect the existence of such caustics (e.g., Onemli \& Sikivie
2007). Compared with sharp drops of DM density in caustics, shocks
in dark matter halo of our model are different in that the
discontinuity of DM density across shocks is less drastic. In
general, the density on the downstream side of shocks are $\sim
2-10$ times of the density on the upstream side of shocks. However,
DM shocks share certain common features with DM caustics. They are
both discontinuous surfaces in dark matter halo and both may be
detected through gravitational lensing effects at least in
principle.

\section*{ACKNOWLEDGEMENT}

This research was supported in part by Tsinghua Centre for
Astrophysics (THCA),
by National Natural Science Foundation of China (NSFC) grants
10373009 and 10533020 and National Basic Science Talent Training
Foundation
(NSFC J0630317) at Tsinghua University, and by the
Yangtze Endowment and the SRFDP 20050003088
at Tsinghua University. Y.-F. Jiang and C.-C. Jin thank W.-G. Wang
and F.-Y. Bian for useful discussion.
\vskip 1.0cm

\begin{appendix}
\section{Self-Similar Nonlinear\\
\ \ \ Ordinary Differential Equations}

On the basis of reduced self-similar equations
$(\ref{da1})-(\ref{v2})$ in the section of Model Formulation, the
reduced radial flow speeds $v_1(x)$ and $v_2(x)$ and the reduced
mass densities $\alpha_1(x)$ and $\alpha_2(x)$ are determined by
four coupled first-order nonlinear ODEs shown below.
\begin{eqnarray}
{d\alpha_1(x)}/{dx}={{\cal A}_1(x)}/{{\cal D}_1(x)}\ ,
\end{eqnarray}
\begin{eqnarray}
{dv_1(x)}/{dx}={{\cal V}_1(x)}/{{\cal D}_1(x)}\ ,
\end{eqnarray}
\begin{eqnarray}
{d\alpha_2(x)}/{dx}={{\cal A}_2(x)}/{{\cal D}_2(x)}\ ,
\end{eqnarray}
\begin{eqnarray}
{dv_2(x)}/{dx}={{\cal V}_2(x)}/{{\cal D}_2(x)}\ .
\end{eqnarray}
The explicit expressions of the two denominators ${\cal D}_1(x)$ and
${\cal D}_2(x)$ are given by
\begin{eqnarray}
{\cal D}_1(x)\equiv
(nx-v_1)^2-\gamma_1\alpha_1^{q_1+\gamma_1-1}x^{2q_1}(nx-v_1)^{q_1}\
, \label{D1}
\end{eqnarray}
and
\begin{eqnarray}
\lefteqn{{\cal D}_2(x)\equiv (\kappa nx-v_2)^2/\kappa}\nonumber\\
&&{}-\kappa^{2q_2-1}\gamma_2\alpha_2^{q_2+\gamma_2-1}x^{2q_2}(\kappa
nx-v_2)^{q_2}\ .\label{D2}
\end{eqnarray}
The explicit expressions for the four numerators ${\cal A}_1(x)$,
${\cal V}_1(x)$, ${\cal A}_2(x)$ and ${\cal V}_2(x)$ are given by
\begin{eqnarray}
\lefteqn{{\cal A}_1(x)\equiv\alpha_1
\bigg\{q_1\alpha_1^{q_1+\gamma_1-1}x^{2q_1}
(nx-v_1)^{q_1-1}(3n-2)}\nonumber\\
&&{}+\frac{\alpha_1}{(3n-2)}(nx-v_1)
+\frac{\alpha_2}{(3n-2)}\bigg(nx-\frac{v_2}{\kappa}\bigg)\nonumber\\
&&{}+(n-1)v_1-\frac{2(x-v_1)(nx-v_1)}{x}\bigg\}\ ;
\end{eqnarray}
\begin{eqnarray}
\lefteqn{{\cal V}_1(x)\equiv q_1\alpha_1^{q_1+\gamma_1-1}x^{2q_1}
(nx-v_1)^{q_1}(3n-2)}\nonumber\\
&&{}-2\gamma_1\alpha_1^{q_1+\gamma_1-1}
x^{2q_1-1}(nx-v_1)^{q_1}(x-v_1)\nonumber\\
&&{}+\frac{\alpha_1(nx-v_1)^2}{(3n-2)}
+\frac{\alpha_2(nx-v_1)}{(3n-2)}
\bigg(nx-\frac{v_2}{\kappa}\bigg)\nonumber\\
&&{}+(n-1)(nx-v_1)v_1\ ;
\end{eqnarray}
\begin{eqnarray}
\lefteqn{{\cal A}_2(x)\equiv \alpha_2
\bigg\{q_2\kappa^{2q_2}\alpha_2^{q_2+\gamma_2-1}
x^{2q_2}(\kappa nx-v_2)^{q_2-1}(3n-2)}\nonumber\\
&&{}+\frac{\alpha_2(\kappa nx-v_2)}{(3n-2)}
+\frac{\kappa\alpha_1(nx-v_1)}{(3n-2)}\nonumber\\
&&{}+(n-1)v_2-\frac{2(\kappa x-v_2)(\kappa nx-v_2)}{\kappa
x}\bigg\}\ ;
\end{eqnarray}
\begin{eqnarray}
\lefteqn{{\cal V}_2(x)\equiv q_2\kappa^{2q_2}
\alpha_2^{q_2+\gamma_2-1}x^{2q_2}
(\kappa nx-v_2)^{q_2}(3n-2)}\nonumber\\
&&-2\kappa^{2q_2-1}\gamma_2\alpha_2^{q_2+\gamma_2-1}
x^{2q_2-1}(\kappa nx-v_2)^{q_2}(\kappa x-v_2)\nonumber\\
&&+\frac{\alpha_2(\kappa nx-v_2)^2}{(3n-2)}
+\frac{\kappa\alpha_1(nx-v_1)}{(3n-2)}(\kappa nx-v_2)\nonumber\\
&&+(n-1)(\kappa nx-v_2)v_2\ .
\end{eqnarray}
Here, all relevant dimensionless parameters are defined in the main
text. With proper asymptotic solutions at large and small $x$ and
shock conditions across the sonic critical curves, we can construct
global semi-complete self-similar solutions with shocks to model
large-scale dynamics in a galaxy cluster involving hot ICM and dark
matter halo.

\section{The Quartic Equation\\
\ \quad\qquad\qquad\qquad for Index Parameter $k$}

In equation (\ref{k}), the index $k$ for the quasi-static solution
of two gravity coupled fluids are determined by the following
quartic equation
\begin{eqnarray}
C_{k,1}\ k^4+C_{k,2}\ k^3+C_{k,3}\ k^2+C_{k,4}\ k+C_{k,5}=0\ ,
\end{eqnarray}
where the five coefficients $C_{k,1},\ C_{k,2},\ C_{k,3},\
C_{k,4},\hbox{ and } C_{k,5}$ are explicitly determined by the
following expressions
\begin{eqnarray}
C_{k,1}\equiv n^4\gamma_1\gamma_2\bigg(1+\frac{A_1}{A_2}\bigg)
\bigg(1+\frac{A_2}{A_1}\bigg)\ ,
\end{eqnarray}
\begin{eqnarray}
C_{k,2}\equiv (6n-8)n^3\gamma_1\gamma_2\bigg(1+\frac{A_1}{A_2}\bigg)
\bigg(1+\frac{A_2}{A_1}\bigg)\ ,
\end{eqnarray}
\begin{eqnarray}
\lefteqn{C_{k,3}\equiv\Big\{-3n^4\gamma_1\gamma_2
+n^2(4-2n)(q_2+\gamma_2-1)}\nonumber\\
&&{}\times[(2-3n)\gamma_1-(4-2n)(q_1-1)]\nonumber\\
&&{}+n^2(4-2n)(q_1+\gamma_1-1)[(2-3n)\gamma_2\nonumber\\
&&{}-(4-2n)(q_2-1)]+(4-2n)^2n^2(q_1-1)(q_2-1)\nonumber\\
&&{}+(4-2n)^2n^2(q_1+\gamma_1-1)(q_2+\gamma_2-1)\Big\}\nonumber\\
&&{}\qquad\qquad\times\bigg(1+\frac{A_1}{A_2}\bigg)
\bigg(1+\frac{A_2}{A_1}\bigg)\nonumber\\
&&{}+(3n-2)(4-2n)n^2\gamma_1\bigg(1+\frac{A_2}{A_1}\bigg)\nonumber\\
&&{}+(3n-2)(4-2n)n^2\gamma_2\bigg(1+\frac{A_1}{A_2}\bigg)\ ,
\end{eqnarray}
\begin{eqnarray}
\lefteqn{C_{k,4}\equiv 2\Big\{[\gamma_2n+(2-n)(q_2-1)]n^3\gamma_1}\nonumber\\
&&{}+[\gamma_1n+(2-n)(q_1-1)]n^3\gamma_2\nonumber\\
&&{}-4[(2-n)(q_1-1)+\gamma_1n][\gamma_2n+(2-n)(q_2-1)]n^2\nonumber\\
&&{}+2n(q_2+\gamma_2-1)(2-n)
[n\gamma_1+(2-n)^2(q_1-1)]\nonumber\\
&&{}+2n(q_1+\gamma_1-1)(2-n)[n\gamma_2+(2-n)^2(q_2-1)]\nonumber\\
&&{}+8(2-n)^2(n-1)n(q_2+\gamma_2-1)(q_1+\gamma_1-1)\Big\}\nonumber\\
&&{}\qquad\times\bigg(1+\frac{A_1}{A_2}\bigg)
\bigg(1+\frac{A_2}{A_1}\bigg)\nonumber\\
&&{}+2(3n-2)(3n-4)(2-n)n\gamma_1\bigg(1+\frac{A_2}{A_1}\bigg)\nonumber\\
&&{}+2(3n-2)(3n-4)(2-n)n\gamma_2\bigg(1+\frac{A_1}{A_2}\bigg)\ ,
\end{eqnarray}
\begin{eqnarray}
\lefteqn{C_{k,5}\equiv 4(3n-2)(2-n)(n^2-6n+4)
\gamma_1\bigg(1+\frac{A_2}{A_1}\bigg)}\nonumber\\
&&{}+4(3n-2)(2-n)(n^2-6n+4)\gamma_2
\bigg(1+\frac{A_1}{A_2}\bigg)\nonumber\\
&&{}+4[\gamma_1n+(2-n)(q_1-1)][\gamma_2n+(2-n)(q_2-1)]\nonumber\\
&&{}\qquad\quad\times n^2\bigg(1+\frac{A_1}{A_2}\bigg)
\bigg(1+\frac{A_2}{A_1}\bigg)\nonumber\\
&&{}+8[\gamma_1n+(2-n)(q_1-1)](2-n)(q_2+\gamma_2-1)\nonumber\\
&&{}\qquad\quad\times n(n-1)\bigg(1+\frac{A_1}{A_2}\bigg)
\bigg(1+\frac{A_2}{A_1}\bigg)\nonumber\\
&&{}+8[\gamma_2n+(2-n)(q_2-1)](2-n)(q_1+\gamma_1-1)\nonumber\\
&&{}\qquad\quad\times n(n-1)\bigg(1+\frac{A_1}{A_2}\bigg)
\bigg(1+\frac{A_2}{A_1}\bigg)\nonumber\\
&&{}+16(n-1)^2(2-n)^2(q_2+\gamma_2-1)\nonumber\\
&&{}\qquad\times (q_1+\gamma_1-1)\bigg(1+\frac{A_1}{A_2}\bigg)
\bigg(1+\frac{A_2}{A_1}\bigg)\nonumber\\
&&{}-4(3n-2)^2(2-n)^2(q_1-1)\bigg(1+\frac{A_2}{A_1}\bigg)\nonumber\\
&&{}-4(3n-2)^2(2-n)^2(q_2-1)\bigg(1+\frac{A_1}{A_2}\bigg)\ ,
\end{eqnarray}
where the two coefficients $A_1$ and $A_2$ in the static SPS
solution (\ref{SPS}) are determined by equation
(\ref{staticdensity}). In short, once the four parameters $(n,\
\gamma_1,\ \gamma_2,\ \kappa)$ are specified, we should be able to
determine the static SPS solution (\ref{SPS}) and the corresponding
quasi-static solution (\ref{quasiv})$-$(\ref{Ni}).

\section{Quasi-static solutions with complex $k$}

For a complex $k$ root with $k=k_1+\hbox{i}k_2$ with $k_1$ and $k_2$
being the real and imaginary parts of $k$ (note that the value of
$k$ is the same for both fluids coupled by gravity), the four
parameters $L_1$, $L_2$, $N_1$ and $N_2$ (the coefficients for
radial flow speeds and mass densities at small $x$ of quasi-static
solutions defined in equation (\ref{quasiv}) and (\ref{quasia})) all
become complex with $L_1=L_{1,1}+\hbox{i}L_{1,2}$,
$L_2=L_{2,1}+\hbox{i}L_{2,2}$, $N_1=N_{1,1}+\hbox{i}N_{1,2}$, and
$N_2=N_{2,1}+\hbox{i}N_{2,2}$, where the real and imaginary parts
are all explicitly written out. As the static SPS solution
(\ref{SPS}) is real, we only take the real part as the quasi-static
solution, namely
\begin{eqnarray}
v_i(x)&=&Re(L_ix^k)=Re[L_ix^{k_1}\exp(ik_2\ln x)]\nonumber\\
&=&x^{k_1}[L_{i,1}\cos(k_2\ln x)-L_{i,2}\sin(k_2\ln x)]\ ,
\end{eqnarray}
\begin{eqnarray}
\alpha_i(x)&=&Re\bigg(A_ix^{-2/n}
+N_ix^{k-1-2/n}\bigg)\nonumber\\
&=&Re\bigg[A_ix^{-2/n}+x^{-2/n-1+k_1}
N_i\exp(ik_2\ln x)\bigg]\nonumber\\
&=&A_ix^{-2/n}+x^{-2/n-1+k_1}\nonumber\\
&&\times[N_{i,1}\cos(k_2\ln x)-N_{i,2}\sin(k_2\ln x)]\ ,
\end{eqnarray}
where
\begin{eqnarray}
\lefteqn{n^2[(k_1-1)^2+k_2^2]N_{1,1}
=(k_1-1)[2(n-1) }\nonumber\\
&&+nk_1]L_{1,1}A_1+(3n-2)k_2L_{1,2}A_1+nk_2^2L_{1,1}A_1\ ,
\end{eqnarray}
\begin{eqnarray}
N_{1,2}&=&N_{1,1}(k_1-1)/k_2+L_{1,2}A_1/n\nonumber\\
&&-[2(n-1)+nk_1]L_{1,1}A_1/(k_2n^2)\ ,
\end{eqnarray}
\begin{eqnarray}
\lefteqn{n^2[(k_1-1)^2+k_2^2]\kappa N_{2,1}
=(k_1-1)[2(n-1) }\nonumber\\
&&{}+nk_1]L_{2,1}A_2+(3n-2)k_2L_{2,2}A_2+nk_2^2L_{2,1}A_2\ ,
\end{eqnarray}
\begin{eqnarray}
N_{2,2}&=&N_{2,1}(k_1-1)/k_2+L_{2,2}A_2/(\kappa n)\nonumber\\
&&-[2(n-1)+nk_1]L_{2,1}A_2/(\kappa k_2n^2)\ .
\end{eqnarray}
As equation (\ref{k}) is a quartic equation for $k$ involving real
coefficients, it has either real roots or pairs of complex conjugate
roots. With free parameters $L_{i,1}$ and $L_{i,2}$, we can always
choose the imaginary part of a complex root\footnote{
Parameter $k_2$ is the imaginary part of a complex $k$ index. If
$k_2<0$, for the $k_2$ that appears in function cos($x$), it does
not matter to choose it to be $-k_2$. For $k_2$ that appears in
function sin($x$), as long as $L_{i,2}$ being a free parameter, we
can always choose it to be $-L_{i,2}$ and it follows that $N_{i,2}$
becomes $-N_{i,2}$. In this manner, solution for $-k_2$ remains the
same as that for $k_2$. In conclusion, there is no loss of
generality to choose $k_2>0$ to represent all possible solutions
with free parameters $L_i$. } $k$ to be $k_2>0$.

For this type of quasi-static solutions with complex $k$, there are
asymptotic oscillations as $x\rightarrow 0^{+}$ (see also Lou \&
Wang 2006 for such oscillatory solution behaviours in a single
fluid).

\section{Asymptotic Series Solutions at Large x}

As stated in subsection \ref{asymptotic solutions at
large x}, for a given value of $n$, the asymptotic series solution
of equations (\ref{da1})$-$(\ref{v2}) at large $x$ is in the form of
\begin{eqnarray}
\alpha_1(x)\rightarrow E_1x^{-2/n}+I_1x^{-3/n}+\dots\ ,
\end{eqnarray}
\begin{eqnarray}
\alpha_2(x)\rightarrow E_2x^{-2/n}+I_2x^{-3/n}+\ldots\ ,
\end{eqnarray}
\begin{eqnarray}
v_1(x)\rightarrow H_1x^{-1/n+1}+G_1x^{-2/n+1}+\ldots\ ,
\end{eqnarray}
\begin{eqnarray}
v_2(x)\rightarrow H_2x^{-1/n+1}+G_2x^{-2/n+1}+\ldots\ .
\end{eqnarray}
The four coefficients $I_1$, $I_2$, $G_1$ and $G_2$ can be expressed
in terms of $E_1$, $E_2$, $H_1$ and $H_2$ as
\begin{eqnarray}
\lefteqn{G_1=\frac{(1-n)}{n}H_1^2+2(2-n)
E_1^{q_1+\gamma_1+1}n^{q_1-1}}\nonumber\\
&&{}-\frac{n}{(3n-2)}(E_1+E_2)\ ,
\end{eqnarray}
\begin{eqnarray}
\lefteqn{G_2=\frac{(1-n)}{n}\frac{H_2^2}{\kappa}+2(2-n)
\kappa^{3q_2-1}E_2^{q_2+\gamma_2+1}n^{q_2-1}}\nonumber\\
&&{}-\frac{\kappa n}{(3n-2)}(E_1+E_2)\ ,
\end{eqnarray}
\begin{eqnarray}
I_1=\frac{3(1-n)}{n}H_1E_1\ ,
\end{eqnarray}
\begin{eqnarray}
I_2=\frac{3(1-n)}{n}\frac{H_2E_2}{\kappa}\ .
\end{eqnarray}
For solutions with $n>1$ and $H_1=H_2=0$, the asymptotic series
solutions finite at large $x$ become
\begin{eqnarray}
\alpha_1(x)\rightarrow E_1x^{-2/n}+F_1x^{-4/n+1}+\ldots\
,\label{largea1_0}
\end{eqnarray}
\begin{eqnarray}
\alpha_2(x)\rightarrow E_2x^{-2/n}+F_2x^{-4/n+1}+\ldots\
,\label{largea2_0}
\end{eqnarray}
\begin{eqnarray}
v_1(x)\rightarrow G_1x^{-2/n+1}+D_1x^{-4/n+2}+\ldots\ ,
\label{largev12}
\end{eqnarray}
\begin{eqnarray}
v_2(x)\rightarrow G_2x^{-2/n+1}+D_2x^{-4/n+2}+\ldots\ .
\label{largev22}
\end{eqnarray}
The six coefficients $G_1$, $G_2$, $F_1$, $F_2$, $D_1$, and $D_2$
are determined by specified values of $E_1$ and $E_2$ below
\begin{eqnarray}
G_1=2(2-n)E_1^{q_1+\gamma_1+1}n^{q_1-1} -\frac{n(E_1+E_2)}{(3n-2)}\
,
\end{eqnarray}
\begin{eqnarray}
G_2=2(2-n)\kappa^{3q_2-1}E_2^{q_2+\gamma_2+1} n^{q_2-1}-\frac{\kappa
n(E_1+E_2)}{(3n-2)}\ ,
\end{eqnarray}
\begin{eqnarray}
\lefteqn{F_1=(4-3n)(2-n)E_1^{q_1+\gamma_1}n^{q_1-2}}\nonumber\\
&&{}\qquad +\frac{(3n-4)}{2(3n-2)}E_1(E_1+E_2)\ ,
\end{eqnarray}
\begin{eqnarray}
\lefteqn{F_2=\kappa^{3q_2-2}(4-3n)(2-n)
E_2^{q_2+\gamma_2}n^{q_2-2}}\nonumber\\
&&{}\qquad +\frac{(3n-4)}{2(3n-2)}E_2(E_1+E_2)\ ,
\end{eqnarray}
\begin{eqnarray}
\lefteqn{D_1=\frac{2}{(3-n)}\bigg[(2-n)(q_1+\gamma_1-1)
E_1^{q_1+\gamma_1-2}n^{q_1-1}F_1}\nonumber\\
&&{}\qquad\qquad\qquad\qquad -\frac{n(F_1+F_2)}{2(3n-2)}\bigg]\ ,
\end{eqnarray}
\begin{eqnarray}
\lefteqn{D_2=\frac{2}{(3-n)}\bigg[(2-n)\kappa^{3q_2-1}
(q_2+\gamma_2-1) }\nonumber\\
&&{} \qquad\quad\times E_2^{q_2+\gamma_2-2}n^{q_2-1}F_2-\frac{\kappa
n(F_1+F_2)}{2(3n-2)}\bigg]\ .
\end{eqnarray}

\section{Eigendirections across the sonic critical curve}

For a specified pair of ($\alpha_2$, $v_2$) at a given
$x$, we now determine the eigendirections across the sonic critical
curve of fluid $1$. The first equation in nonlinear ODEs (Equation
(\ref{da1})) gives
\begin{eqnarray}
\frac{d\alpha_1}{dx}=\bigg[\alpha_1\frac{dv_1}{dx}
-2\frac{(x-v_1)}{x}\alpha_1\bigg]\bigg/(nx-v_1)\ .
\end{eqnarray}
We take $n>2/3$ and thus $nx-v>0$ in this paper. From ${\cal
D}_1(x)=0$, we obtain
\begin{eqnarray}
v_1=nx-\Big[\gamma_1x^{2q_1}\alpha_1^{q_1+\gamma_1-1}\Big]^{1/(2-q_1)}\
.
\end{eqnarray}
From ${\cal A}_1(x)=0$,  the value of $\alpha_1$ on the sonic
critical curve at $x$ is determined by the following equation
\begin{eqnarray}
\lefteqn{\bigg(\frac{\alpha_1}{3n-2}+n+1+\frac{2n-4}{\gamma_1}\bigg)
(\gamma_1\alpha_1^{q_1+\gamma_1-1}x^{2q_1})^{1/(2-q_1)}}\nonumber\\
&&{}\quad -2\Big(\gamma_1\alpha_1^{q_1+\gamma_1-1}\Big)^{2/(2-q_1)}
x^{(5q_1-2)/(2-q_1)}\nonumber\\
&&{}=-\frac{\alpha_2}{(3n-2)}\bigg(nx-\frac{v_2}{\kappa}\bigg)-n(n-1)x\
.
\end{eqnarray}
In general, fluid $2$ is not on the sonic critical curve at this
position $x$. Therefore, $\alpha_2^\prime$ is determined by equation
(\ref{da2}) while $v_2^\prime$ is determined by equation (\ref{v2}).
We now apply the L$^\prime$H$\hat{\hbox{o}}$spital rule to determine
the eigenvalues of $v_1^{\prime}(x)$ from equation (\ref{dv1})
\begin{eqnarray}
\frac{dv_1(x)}{dx}=\frac{{\cal V}_1^{'}(x)}{{\cal D}_1^{'}(x)}\ .
\end{eqnarray}
This equation appears to be a quadratic algebraic equation in terms
of $v_1^\prime(x)$, namely
\begin{eqnarray}
C_{v,1}\ [v_1^\prime(x)]^2+C_{v,2}\ v_1^\prime(x)+C_{v,3}=0\ ,
\end{eqnarray}
where the three coefficients $C_{v,1},\ C_{v,2},\ C_{v,3}$ are
explicitly defined by
\begin{eqnarray}\label{eigenv1}
C_{v,1}\equiv
1+\alpha_1^{q_1+\gamma_1-1}\gamma_1^2x^{2q_1}(nx-v_1)^{q_1-2}\ ,
\end{eqnarray}
\begin{eqnarray}
C_{v,2}&\equiv\alpha_1^{q_1+\gamma_1-1}\gamma_1x^{2q_1-1}
(nx-v_1)^{q_1-2}\{4(\gamma_1-1)v_1\nonumber\\
&+[3n(q_1+1)-2(q_1+\gamma_1+1)]x\}-1\ ,
\end{eqnarray}
\begin{eqnarray}
C_{v,3}&\equiv &2\alpha_1^{q_1+\gamma_1-1}x^{2q_1-2}(nx-v_1)^{q_1-2}
\bigg\{\gamma_1(2\gamma_1-1)v_1^2\nonumber\\
&+&\Big[n(3q_1\gamma_1-2)-2[\gamma_1(q_1+\gamma_1+1)-2]\Big]xv_1\nonumber\\
&-&(n-2)[n-3nq_1+2(q_1+\gamma_1-1)]x^2\bigg\}
\nonumber\\
&+&\frac{n\alpha_2}{(3n-2)}+ \frac{\alpha_1(n-2+2v_1/x)}{(3n-2)}
\nonumber\\
&-&\frac{\alpha_2v_2^\prime}{(3n-2)\kappa }
-\frac{\alpha_2^\prime
v_2}{(3n-2)\kappa}+\frac{nx\alpha_2^\prime}{(3n-2)}\ .
\end{eqnarray}
We can then readily solve for the two eigenvalues of
$v_1^{\prime}(x)$, corresponding two eigendirections. The two
possible eigendirections across the corresponding sonic critical
curve of fluid $2$ can be computed in the same manner.

\section{Shock Conditions}

As stated in subsection of self-similar shocks
\ref{self-similar shock}, all variables with subscript $\{u,i\}$
refer to variables on the upstream side of a shock for fluid $i$,
while all variables with subscript $\{d,i\}$ refer to variables on
the downstream side of a shock for fluid $i$. Convenient variables
$\Gamma_{u,i}$ and $\Gamma_{d,i}$ are defined by
\begin{eqnarray}
\Gamma_{d,i}\equiv n-\frac{v_{d,i}}{x_{d,i}}\ ,
\end{eqnarray}
\begin{eqnarray}
\Gamma_{u,i}\equiv n-\frac{v_{u,i}}{x_{u,i}}\ .
\end{eqnarray}
Parameter $\lambda_i$ defined by
\begin{eqnarray}
\lambda_i=\bigg(\frac{K_{d,i}}{K_{u,i}}\bigg)^{1/2}\
\end{eqnarray}
is essentially the sound speed ratio of downstream and upstream
sides of a shock and should be greater than $1$ for entropy increase
on the downstream side. According to the dimensionless equation of
state (see equation (\ref{stateless})), the dimensional equation of
state is
\begin{eqnarray}
P_i=K_i^{(1-3q_i/2)}(4\pi)^{\gamma_i-1}G^{\gamma_i-1+q_i}
(3n-2)^{q_i}M_i^{q_i}\rho_i^{\gamma_i}\ .
\end{eqnarray}
Then shock equations (equation of mass conversation, radial momentum
conservation equation and energy conservation equation
(\ref{shockmass})$-$(\ref{shockenergy})) can be rearranged into
dimensionless self-similar forms of
\begin{eqnarray}
\lefteqn{\alpha_{u,i}(nx_{u,i}-v_{u,i})
=\alpha_{d,i}(nx_{u,i}-\lambda_iv_{d,i})}\nonumber\\
&&{}=\alpha_{d,i}\lambda_i(nx_{d,i}-v_{d,i})\ ,
\end{eqnarray}
\begin{eqnarray}
\lefteqn{\lambda^2_i\Big[\alpha_{d,i}^{q_i+\gamma_i}
x_{d,i}^{2q_i}(nx_{d,i}-v_{d,i})^{q_i}
+\alpha_{d,i}(nx_{d,i}-v_{d,i})^2\Big]}\nonumber\\
&=\alpha_{u,i}^{q_i+\gamma_i}x_{u,i}^{2q_i}
(nx_{u,i}-v_{u,i})^{q_i}+\alpha_{u,i}(nx_{u,i}-v_{u,i})^2\ ,
\end{eqnarray}
\begin{eqnarray}
\lefteqn{\lambda_i^2\bigg[\frac{(nx_{d,i}-v_{d,i})^2}{2}
+\frac{\gamma_i\alpha_{d,i}^{q_i+\gamma_i-1}}{(\gamma_i-1)}
x_{d,i}^{2q_i}(nx_{d,i}-v_{d,i})^{q_i}\bigg]}\nonumber\\
&=(nx_{u,i}-v_{u,i})^2/2\qquad\qquad\qquad\qquad\qquad\nonumber\\
&\qquad\quad +\gamma_i \alpha_{u,i}^{q_i+\gamma_i-1}
x_{u,i}^{2q_i}(nx_{u,i}-v_{u,i})^{q_i}/(\gamma_i-1)\ .
\end{eqnarray}
With $\Gamma_1$ and $\Gamma_2$ for the two coupled fluids, we can
rewrite the above three equations in more symmetric forms, namely
\begin{eqnarray}\label{eq:86}
\alpha_{u,i}\Gamma_{u,i}=\alpha_{d,i}\Gamma_{d,i}\ ,
\end{eqnarray}
\begin{eqnarray}\label{eq:87}
\lefteqn{\alpha_{d,i}^{q_i+\gamma_i}
x_{d,i}^{3q_i-2}\Gamma_{d,i}^{q_i}
+\alpha_{d,i}\Gamma_{d,i}^2}\nonumber\\
&&{}=\alpha_{u,i}^{q_i+\gamma_i} x_{u,i}^{3q_i-2}\Gamma_{u,i}^{q_i}
+\alpha_{u,i}\Gamma_{u,i}^2\ ,
\end{eqnarray}
\begin{eqnarray}\label{eq:88}
\lefteqn{\frac{\Gamma_{d,i}^2}{2}+\frac{\gamma_i}
{(\gamma_i-1)}\alpha_{d,i}^{q_i+\gamma_i-1}
x_{d,i}^{3q_i-2}\Gamma_{d,i}^{q_i}}\nonumber\\
&&{}=\frac{\Gamma_{u,i}^2}{2}+\frac{\gamma_i}
{(\gamma_i-1)}\alpha_{u,i}^{q_i+\gamma_i-1}
x_{u,i}^{3q_i-2}\Gamma_{u,i}^{q_i}\ .
\end{eqnarray}
From above three equations (\ref{eq:86})$-$(\ref{eq:88}),
$\Gamma_{u,i}$ upstream of a shock can be determined by the
variables on the downstream side of a shock from the following
quadratic equation
\begin{eqnarray}
&&\frac{(\gamma_i+1)}{2\gamma_i}\Gamma_{u,i}^2
-\Big(\alpha_{d,i}^{q_i+\gamma_i-1}
\Gamma_{d,i}^{q_i-1}x_{d,i}^{3q_i-2}
+\Gamma_{d,i}\Big)\Gamma_{u,i}\qquad\qquad\nonumber\\
&&\qquad\quad+\alpha_{d,i}^{q_i+\gamma_i-1}
\Gamma_{d,i}^{q_i}x_{d,i}^{3q_i-2}
+\frac{(\gamma_i-1)}{2\gamma_i}\Gamma_{d,i}^2=0\ ,\label{Gamma}
\end{eqnarray}
where subscript $i=1,\ 2$ correspond to fluid 1 and 2, respectively.
One root of quadratic equation (\ref{Gamma}) is a trivial solution
$\Gamma_{u,i}=\Gamma_{d,i}$, which is omitted. The other root of
physical relevance is given in the subsection of self-similar shocks
\ref{self-similar shock}.

\section{Sound Speed Ratio}

In our two-fluid model framework, the velocity dispersion of dark
matter particles (DMPs) is mimicked by an equivalent `sound speed'.
This velocity dispersion of DMPs produces a pressure-like effect. In
this scenario, the polytropic sound speed for fluid $i$ is given by
\begin{eqnarray}
a_{i}=\bigg(\frac{\partial{P_{i}}}
{\partial{\rho_{i}}}\bigg)_s^{1/2}=\bigg(\frac{\gamma_i
P_{i}}{\rho_{i}}\bigg)^{1/2}\ .\label{sound}
\end{eqnarray}
By self-similar transformation (\ref{transformation}), this
polytropic sound speed can be written as
\begin{eqnarray}
a_i=\Big(\gamma_i\frac{\beta_i}{\alpha_i}\Big)^{1/2}
K_i^{1/2}t^{n-1}\ .
\end{eqnarray}
As the reduced pressure $\beta_i$ is determined by the reduced
equation of state (\ref{stateless}) and there is an expression
(\ref{mass}) for the reduced enclosed mass, the polytropic sound
speed in fluid $i$ can be written as
\begin{eqnarray}
a_i=\gamma_i^{1/2}K_i^{1/2}t^{n-1}\alpha_i^{(q_i+\gamma_i-1)/2}
x_i^{q_i}(nx_i-v_i)^{q_i/2}\ .
\end{eqnarray}
Since $\big({K_1}/{K_2}\big)^{1/2}={x_2}/{x_1}=\kappa$, the sound
speed ratio of fluid $1$ to fluid $2$ is given by
\begin{eqnarray}
\frac{a_1}{a_2}=\Big(\frac{\gamma_1}{\gamma_2}\Big)^{1/2}\kappa^{1-q_2}\
\frac{\alpha_1^{(q_1+\gamma_1-1)/2}}{\alpha_2^{(q_2+\gamma_2-1)/2}}\
x^{q_1-q_2}\frac{(nx_1-v_1)^{q_1/2}}{(nx_2-v_2)^{q_2/2}}\
.\label{soundratio}
\end{eqnarray}
The sound speed varies with radius and the sound speed ratio at
different radii can be computed according to equation
(\ref{soundratio}).

In the regime of $x\rightarrow 0^{+}$, we take the quasi-static
solution of our model and the asymptotic behaviour of density is
determined by equation (\ref{quasia}) (velocities are small
compared with $x$). Thus when $x\rightarrow 0^{+}$, the sound
speed ratio has the asymptotic behaviour
\begin{eqnarray}
\frac{a_1}{a_2}\rightarrow\Big(\frac{\gamma_1}
{\gamma_2}\Big)^{1/2}\kappa^{1-3q_2/2}\
\frac{A_1^{(q_1+\gamma_1-1)/2}}{A_2^{(q_2+\gamma_2-1)/2}}\ .
\end{eqnarray}
As the coefficients of static polytropic solution $A_1$ and $A_2$
are determined by equation (\ref{staticdensity}), the sound speed
ratio at small radius is
\begin{eqnarray}
\frac{a_1}{a_2}=\Big(\frac{\gamma_1\ n^{q_2}}{\gamma_2\
n^{q_1}}\Big)^{1/2}\ .
\end{eqnarray}
Therefore in the limit of $x\rightarrow 0^{+}$ and for the
quasi-static solution, the sound speed ratio is independent of
$\kappa$ and is only dependent on $n$ and on the two polytropic
indices $\gamma_i$.

In the limit of $x\rightarrow+\infty$, the asymptotic behaviour of
density is determined by equation (\ref{largea1}) and equation
(\ref{largea2}) (velocities are small compared with $x$), so the
asymptotic behaviour of sound speed ratio at large $x$ is
\begin{eqnarray}
\frac{a_1}{a_2}\rightarrow\Big(\frac{\gamma_1}
{\gamma_2}\Big)^{1/2}\kappa^{1-3q_2/2}\
\frac{E_1^{(q_1+\gamma_1-1)/2}}{E_2^{(q_2+\gamma_2-1)/2}}\ ,
\end{eqnarray}
where $E_1$ and $E_2$ are two coefficients determined by equations
(\ref{largea1}) and (\ref{largea2}).

Two specific examples for the radial profile of polytropic sound
speed ratio $a_2/a_1$ are shown in Fig. \ref{nsmall1_soundratio} and
Fig. \ref{neq1_soundratio} in the main text.
\end{appendix}

\end{document}